\documentclass[aps,prb,twocolumn,superscriptaddress,10pt]{revtex4-1}

\usepackage{amsmath}
\usepackage{amssymb}
\usepackage{color}
\usepackage{enumitem}
\usepackage{hyperref}
\usepackage{microtype}
\usepackage[percent]{overpic}
\usepackage{verbatim}

\newcommand{\R}{\mathbb{R}}
\newcommand{\p}{\partial}

\definecolor{webgreen}{rgb}{0,.35,0}
\definecolor{webbrown}{rgb}{.6,0,0}
\definecolor{royalblue}{rgb}{0,0,0.9}

\hypersetup{
   colorlinks=true, linktocpage=true, pdfstartpage=3, pdfstartview=FitV,
   breaklinks=true, pdfpagemode=UseNone, pageanchor=true, pdfpagemode=UseOutlines,
   plainpages=false, bookmarksnumbered, bookmarksopen=true, bookmarksopenlevel=1,
   hypertexnames=true, pdfhighlight=/O,
   urlcolor=webbrown, linkcolor=royalblue, citecolor=webgreen,
   pdftitle={Voronoi cell analysis: The shapes of particle systems},
   pdfauthor={Emanuel A. Lazar, Jiayin Lu, and Chris H. Rycroft},
   pdfsubject={Voronoi cell analysis, computational physics},
   pdfkeywords={Voronoi tessellations, computational physics, applied physics, computational geometry, numerical analysis},
   pdfcreator={pdfLaTeX},
   pdfproducer={LaTeX with hyperref}
}

\begin{document}

\title{Voronoi cell analysis: The shapes of particle systems}

\author{Emanuel A. Lazar}
\email{mlazar@math.biu.ac.il}
\affiliation{Department of Mathematics, Bar-Ilan University, Ramat Gan 5290002, Israel}

\author{Jiayin Lu}
\email{jiayin\_\,lu@g.harvard.edu}
\affiliation{John A. Paulson School of Engineering and Applied Sciences, Harvard University, 29 Oxford Street, Cambridge, Massachusetts 02138, USA}

\author{Chris H. Rycroft}
\email{chr@seas.harvard.edu}
\affiliation{John A. Paulson School of Engineering and Applied Sciences, Harvard University, 29 Oxford Street, Cambridge, Massachusetts 02138, USA}
\affiliation{Mathematics Group, Lawrence Berkeley National Laboratory, 1 Cyclotron Road, Berkeley, California 94720, USA}

\date{\today}

\begin{abstract}
Many physical systems can be studied as collections of particles embedded in space, often evolving in time.
Natural questions arise concerning how to characterize these arrangements -- are they ordered or disordered?
If they are ordered, how are they ordered and what kinds of defects do they possess?
Voronoi tessellations, originally introduced to study problems in pure mathematics, have become a powerful and versatile tool for analyzing countless problems in pure and applied physics.
We explain the basics of Voronoi tessellations and the shapes they produce, and describe how they can be used to characterize many physical systems.
\end{abstract}

\maketitle

\section{Introduction}

Physical systems can often be abstracted as large sets of point-like particles in space, with each point representing, for example, the position of an atom, colloidal particle, organism, or celestial body.
Fundamental questions arise when studying these systems regarding the manner in which their constituent particles are organized.
Are they ordered or disordered?
If they are ordered, are the particles arranged in a crystal-like fashion, and if so, what kind?
Even systems that are primarily ordered often contain defects of varying dimension and kind, all of which can impact the macroscopic static and dynamic properties of a system.
How can these defects be classified?
In systems that are nominally disordered, such as liquids, glasses, and granular materials, how can the disorder be described in a practical and effective manner?

Voronoi tessellations provide a natural tool for translating questions about the arrangements of particles into ones about polygons and polyhedra, possibly irregular, and about how they fit together to tile space.
The analysis of these shapes can tell us much about particle-like systems on a wide range of scales.

\section{Voronoi cells}

Given a discrete set of particles, the \textit{Voronoi cell} of each particle is the region of space closer to it than to any other particle.\cite{voronoi1908nouvelles}  The term \textit{Voronoi tessellation} is typically used to denote the collection of all Voronoi cells of a set of particles.
Formally, given a metric space $(M, d)$ and a discrete set of particle positions $\{s_1, s_2,\ldots\} \subset M$, the Voronoi cell of a particle $s_i$ is the set
\begin{equation}
V(s_i) = \{x \in M \, | \, d(x,s_i) \leq d(x, s_j), \quad\forall\,\, i\neq j\}.
\label{eqn:voronoi}
\end{equation}
In many applications, the space $M$ under consideration is $\R^2$ or $\R^3$, or some subregion thereof, and the metric $d$ is the standard Euclidean one.  In two dimensions, Voronoi cells are convex polygons that together partition space, while in three dimensions they are convex polyhedra.

\setlength{\fboxsep}{0pt}
\setlength{\unitlength}{0.001\textwidth}
\begin{figure}
  \begin{center}
    \begin{picture}(474,225)
      \put(28,5){\includegraphics[width=0.21\textwidth]{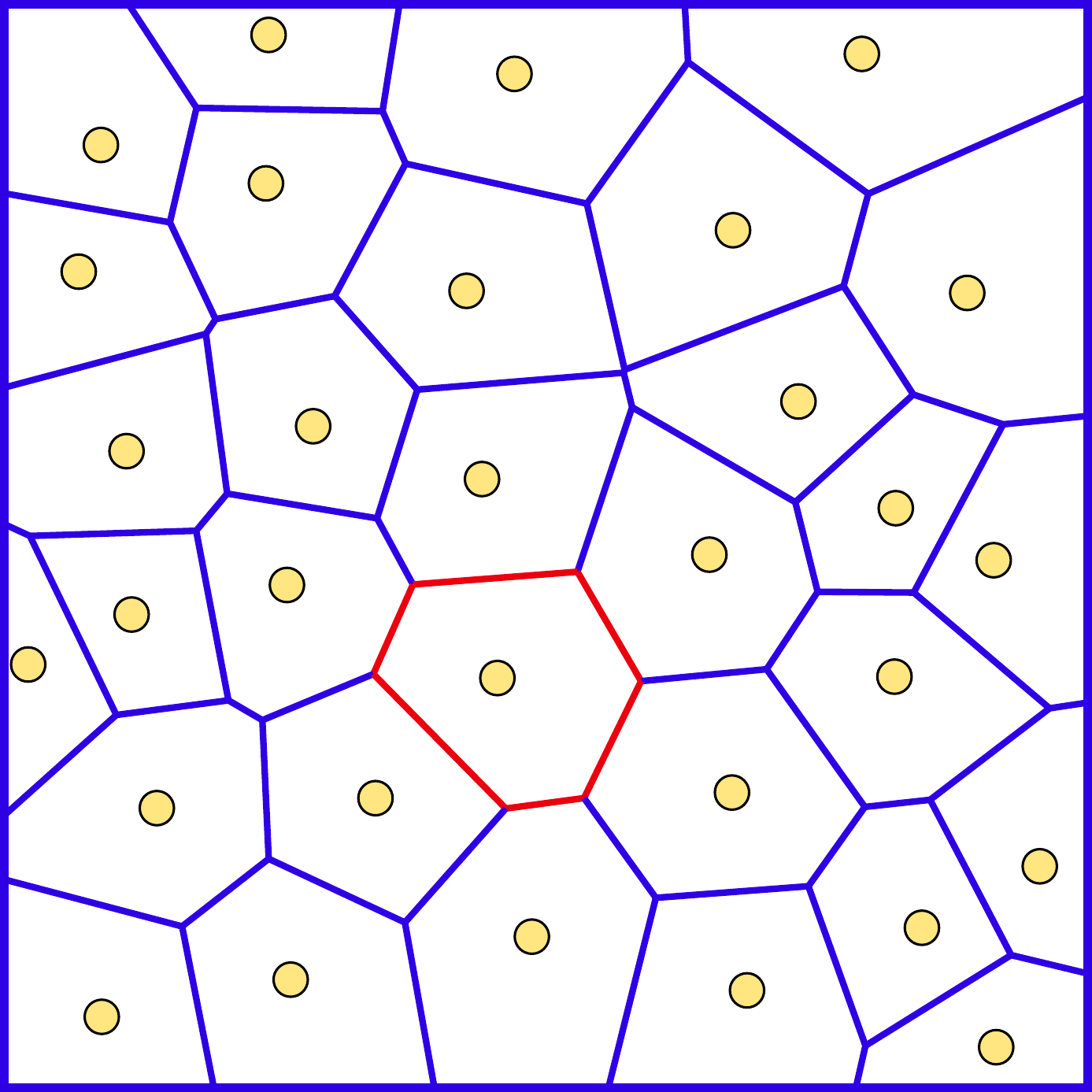}}
      \put(263,0){\includegraphics[width=0.222\textwidth]{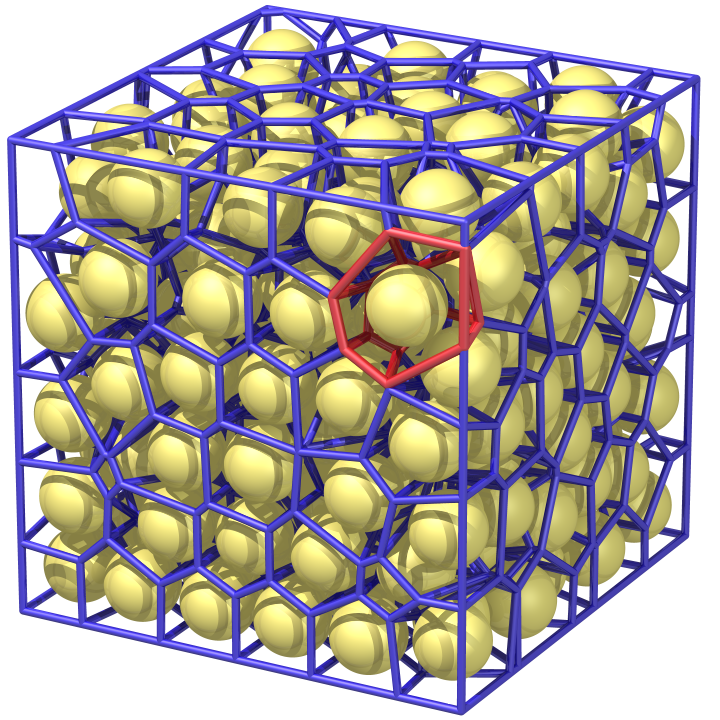}}
      \put(0,200){(a)}
      \put(263,200){(b)}
    \end{picture}
  \end{center}\vspace{-0.5em}
  \caption{(a) Voronoi tessellation in two dimensions. For a set of particles shown by the yellow circles, the blue network divides the space into polygonal regions called Voronoi cells, each of which contains the area closer to one particle than to any other. The outline of one Voronoi cell is shown in red. (b) Three-dimensional Voronoi tessellation for a set of particles shown as yellow spheres. The blue lines show the entire Voronoi tessellation, while the red lines highlight a single Voronoi cell as a convex polyhedron.\label{fig:tessellations}}
\end{figure}

\begin{figure*}
\begin{center}
\begin{overpic}[width=\textwidth]{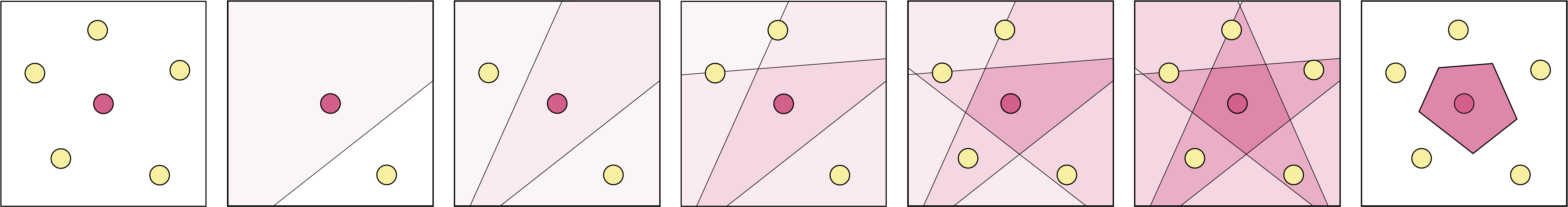}
\put (0.25,11.6) {\footnotesize (a)}
\put (14.75,11.6) {\footnotesize (b)}
\put (29.25,11.6) {\footnotesize (c)}
\put (43.75,11.6) {\footnotesize (d)}
\put (58.25,11.6) {\footnotesize (e)}
\put (72.75,11.6) {\footnotesize (f)}
\put (87.15,11.6) {\footnotesize (g)}
\end{overpic}
\caption{(a) A central magenta particle and five neighboring yellow particles; (b)--(f) construction of the Voronoi cell of the central particle, shown as a series of intersections of half-spaces, where each is defined by the central particle and one of its neighbors. (g) The completed Voronoi cell of the central particle. \label{fig:construction2d}}
\end{center}
\end{figure*}
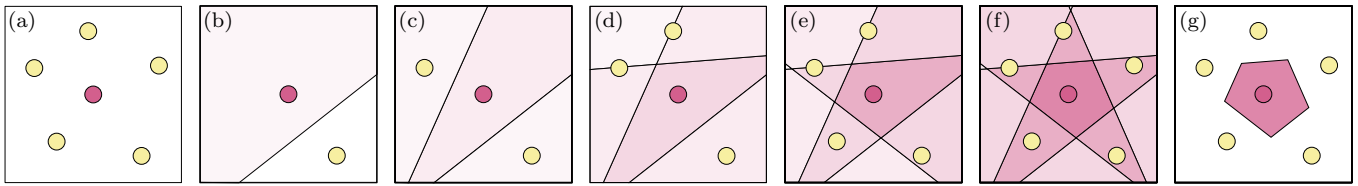

Figure \ref{fig:tessellations} shows examples of particles and their Voronoi cells in two- and three-dimensional systems. The Voronoi cells have different areas and volumes, as well as different numbers of edges and faces. The shape of each Voronoi cell describes the order in the vicinity of a particle, and in the system more generally. In perfect crystals, for example, all Voronoi cells are identical, whereas in disordered ones there can be significant variation of their properties.

Voronoi cells also provide a scale-free mechanism to characterize adjacencies
between particles: two particles are defined to be neighbors if and only if
they share a Voronoi edge (in $\R^2$) or face (in $\R^3$). If edges are drawn
between all such neighbors, this results in the Delaunay triangulation,
another widely-used construction in computational geometry\cite{loera} (see Sec.~\ref{sub:num_an}). 
Voronoi tessellations and Delaunay triangulations
are sometimes referred to as \textit{dual} objects,
because each can be used to construct the other.
Decomposing a region into polyhedral tiles through the Voronoi tessellation
has proven to be a powerful
tool in many branches of science and engineering.\cite{1992okabe}

In some cases, the particles being analyzed may have different radii $r_i$. A
natural method to handle this case is to generalize the definition in
Eq.~\eqref{eqn:voronoi}:
\begin{equation}
  V(s_i) = \{x \in M \, | \, d(x,s_i)-r_i \leq d(x, s_j)-r_j, \,\,\forall\,\, i\neq j\}.
  \label{eqn:s_voronoi}
\end{equation}
This definition results in an alternative tessellation referred to as the \textit{Voronoi S cell};
if all $r_i$ are identical, these cells are equivalent to
those defined by Eq.~\ref{eqn:voronoi}. Although it is possible to work
directly with S cells,\cite{medvedev06,pinheiro13a} they are
much more difficult to compute and analyze, because they typically have curved,
hyperboloidal faces and are thus no longer convex polyhedra. 
For this reason, it is
useful to consider a different generalization,
\begin{equation}
  V(s_i) = \{x \in M \, | \, d(x,s_i)^2-r_i^2 \leq d(x, s_j)^2-r_j^2, \quad\forall\,\, i\neq j\},
  \label{eqn:rad_voronoi}
\end{equation}
which is referred to as the \textit{radical Voronoi tessellation}. In Euclidean 
space, the radical Voronoi tessellation results in convex polyhedra, which 
well-approximate many features of Voronoi S cells\cite{phillips10} at substantially lower computational cost. They are
therefore a popular choice in analyzing systems of particles with unequal
radii.\cite{marck96,sastry97,yi12} In the remainder of this article we focus on
the original Voronoi tessellation given in Eq.~\eqref{eqn:voronoi}, 
although many of the results generalize to the radical Voronoi tessellation
via an adjustment of the positions of the planar Voronoi faces.

\subsection{Construction}
\label{sec:construct}

The definition given by Eq.~\eqref{eqn:voronoi} does not directly lend itself to a practical algorithm for constructing Voronoi cells.
It is thus productive to also think about Voronoi cells from an alternative perspective.
Consider the two particles in Fig.~\ref{fig:construction2d}(b) and the line separating them.
On one side of the line lies the region of space closer to the magenta particle, and on the other side lies the region of space closer to the yellow one; points on the line itself are equidistant to the two particles. If these were the only particles in the system, we would be left with two Voronoi cells, one on each side of the dividing line.
Each of these Voronoi cells is called a \textit{half-space}.
In three-dimensional spaces, the two Voronoi cells are separated by a plane instead of a line.

Imagine now adding a third particle to the system, as in Fig.~\ref{fig:construction2d}(c).
The set of points closer to the magenta particle than to the new yellow particle is again a half-space, and is bounded by a separate line.
Because the Voronoi cell of the magenta particle consists of all points closer to it than to \textit{any} of the other particles, it can be thought of as the intersection of these two half-spaces. Formally, we use
\begin{equation}
H_{ij} = \{x \in M \, | \, d(x,s_i) \leq d(x, s_j)\}
\label{eqn:voronoi2a}
\end{equation}
to denote the half-space of all points as close or closer to $s_i$ than to $s_j$, and then define the Voronoi cell of $s_i$ as the intersection of all such half-spaces,
\begin{equation}
V(s_i) = \bigcap_j H_{ij},
\label{eqn:voronoi2b}
\end{equation}
where $j$ indexes all other particles in the system.  Figures~\ref{fig:construction2d}(d)--(f) contain points at which multiple lines intersect; such points are equidistant to three particles.  Figure~\ref{fig:construction2d}(g) shows the final Voronoi cell, constructed by considering  the central particle and the neighboring yellow ones; it is the intersection of five half-spaces.

Particles located far away from a central particle do not influence its Voronoi cell, because associated half-spaces completely contain the intersection of half-spaces associated to closer particles.  The shape of a Voronoi cell is thus generally determined only by nearby particles.

The definition provided by Eqs.~\eqref{eqn:voronoi2a} and \eqref{eqn:voronoi2b} thus motivates a simple constructive algorithm for building Voronoi cells: iteratively cut the region around a particle $s_i$ by intersecting half-spaces.  Figure~\ref{fig:cell_ov} illustrates the cutting of a three-dimensional Voronoi cell through a half-space associated with an added particle.

\setlength{\unitlength}{0.01\linewidth}
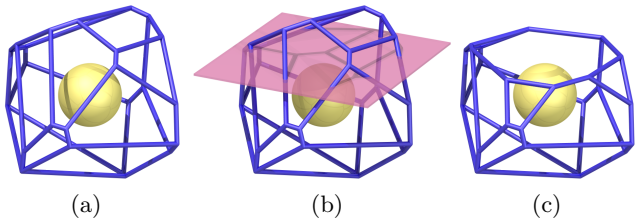
\begin{figure}
  \begin{center}
    \begin{picture}(100,33)
      \put(0,3){\includegraphics[height=0.3\linewidth]{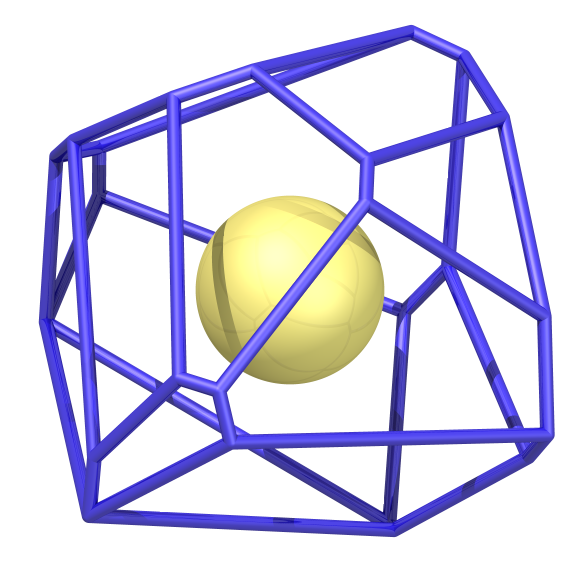}}
      \put(12,0){(a)}
      \put(49,0){(b)}
      \put(70,3){\includegraphics[height=0.3\linewidth]{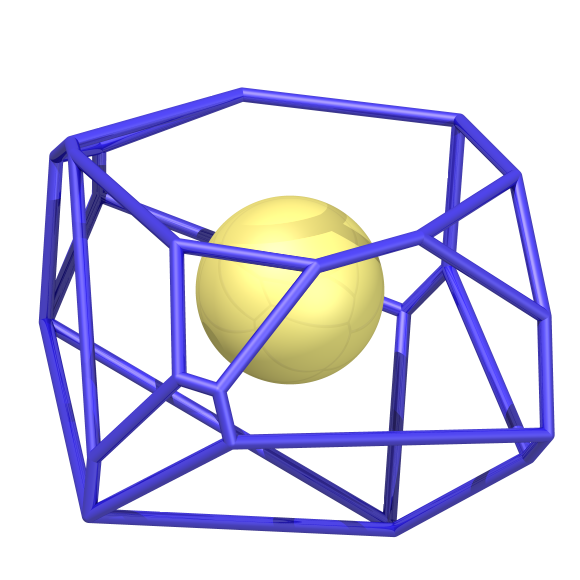}}
      \put(30,3){\includegraphics[height=0.3\linewidth]{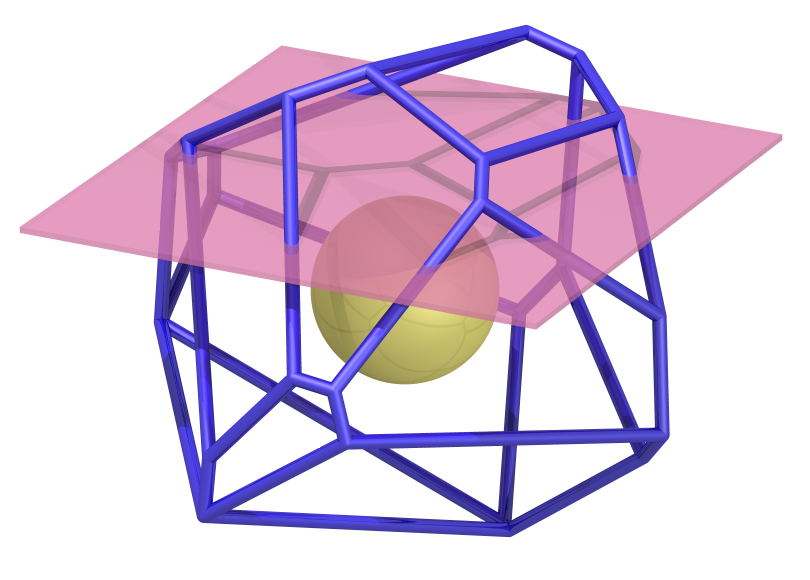}}
      \put(83,0){(c)}
    \end{picture}
  \end{center}\vspace{-1.5em}
  \caption{(a) A central particle and its Voronoi cell; the cell can be represented as a collection of vertices and edges that represent an arbitrary irregular polyhedron.
  (b) Consideration of an additional neighboring particle results in intersecting the Voronoi cell with an additional half-space, whose boundary is shown in magenta. (c) The updated Voronoi
  cell.\label{fig:cell_ov}}
\end{figure}

\subsection{Algorithmic and computational aspects}

The time and memory necessary to compute and represent a Voronoi tessellation of an entire system increases with the number of particles. In a system of $n$ particles, constructing the Voronoi cell of each particle requires, in theory, computing the intersection of $n-1$ half-spaces, and so calculating all Voronoi cells would require consideration of $O(n^2)$ half-spaces.\cite{aurenhammer1991voronoi} However, as described in Sec.~\ref{sec:construct}, usually only the half-spaces for nearby particles influence the Voronoi cell. This consideration allows efficient algorithms to be designed\cite{bentley80} that first initialize the candidate Voronoi cell to fill the entire domain, and then sweep outward, performing half-space intersections for particles of increasing separation, until a termination criterion is reached and it is possible to guarantee that any particles further away cannot possibly affect the Voronoi cell shape. For example, if the candidate Voronoi cell is contained within a sphere $S$ of radius $R$, and all particles within the sphere of radius $2R$ have been considered, then we can guarantee that the Voronoi cell computation is complete, because the half-space for any particle further away will completely contain $S$.

For typical particle arrangements that are evenly distributed, each Voronoi cell requires $O(1)$ half-space intersections, meaning that the total computation time scales linearly with the number of particles. However, in certain pathological cases, the computation time may not be linear.  For example, in three dimensions define $L=\{-n,-(n-1),\ldots, n\}$, and consider the two sets of particles $P_1 = \{(1,k,0) \,|\, k \in L\}$ and $P_2 = \{(-1,0,k) \, | \, k \in L\}$. Then the Voronoi cell for a particle in $P_1$ contains a face for every particle in $P_2$ and vice versa, and thus each Voronoi cell requires $O(n)$ half-space intersections in this case, meaning that the total computation time will scale at least quadratically.

In two dimensions, Voronoi cells can also be constructed using a ``walking method.''\cite{brassel79,cromley85} This approach traces around the edges and vertices of the Voronoi cell, directly constructing it without explicitly considering the half-spaces. Efficient implementations allow this procedure to be performed in $O(1)$ time for a typical Voronoi cell.\cite{maus84} However, unlike half-space intersections, this method does not have an obvious generalization to higher dimensions.

\begin{figure}[b]
  \begin{center}
    \begin{picture}(80,35)
      \put(5,0){\includegraphics[width=0.76\linewidth]{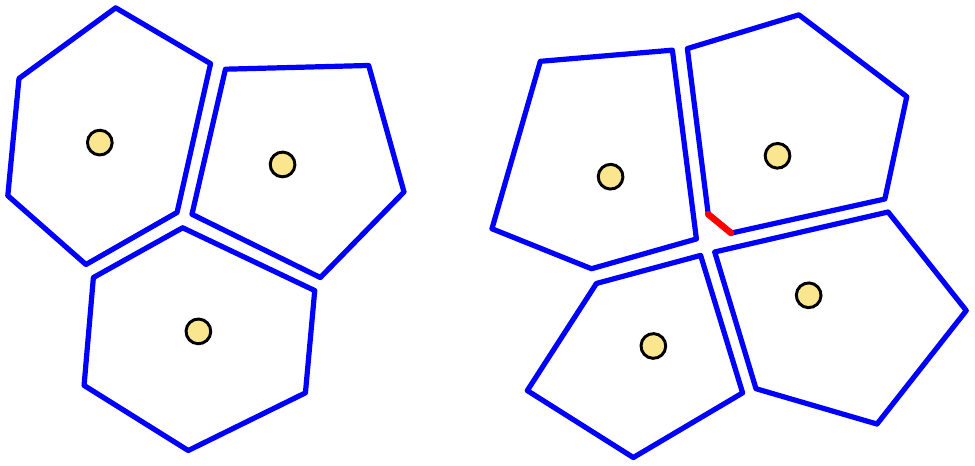}}
      \put(0,30){(a)}
      \put(40,30){(b)}
    \end{picture}
  \end{center}\vspace{-1.4em}
  \caption{(a) Typical case where three Voronoi cells (blue polygons) for three
  particles (yellow circles) meet at a vertex in two dimensions. (b) Special
  case where four Voronoi cells meet at a vertex that is equidistant from
  four particles; floating-point errors could lead to additional small edges (red)
  for some cells. (The Voronoi cells would normally touch but are spaced
  slightly apart from each other for illustrative purposes.)\label{fig:cell_degen}}
\end{figure}

Because each Voronoi cell can be computed and analyzed independently, the cell-based perspective is well-suited to current computational hardware, where parallelism is emphasized. A typical workflow could involve calculating a Voronoi cell, storing its statistics, and then deleting it and moving onto the next. On a multi-core or parallel computer, each thread or process can compute subsets of Voronoi cells independently of the rest. A drawback of the cell-based approach is that it lacks explicit topological information about how the cells are connected together---it is not straightforward to identify common faces, edges, and vertices between cells. This can be a problem once the effects of numerical rounding error of floating point arithmetic~\cite{heath} are accounted for. In a typical situation in two dimensions, three Voronoi cells will meet at a single vertex [see Fig.~\ref{fig:cell_degen}(a)]. However, if the particles are precisely aligned, then it is possible that four or more cells meet at one vertex. In this case, due to numerical rounding error, one or more Voronoi cells may have additional small faces, leading to inconsistent topology between the cells [see Fig.~\ref{fig:cell_degen}(b)]. This problem is not unique to $\R^2$ and happens in arbitrary dimensions. In $\R^n$, it occurs when $n+1$ precisely aligned particles are equidistant to a single point.

\setlength{\fboxsep}{0pt}
\setlength{\tabcolsep}{0.2em}
\begin{figure*}
\begin{center}
\begin{tabular}{ccccc}
\fbox{\includegraphics[trim={230mm 230mm 230mm 230mm},clip,height=0.42\columnwidth]{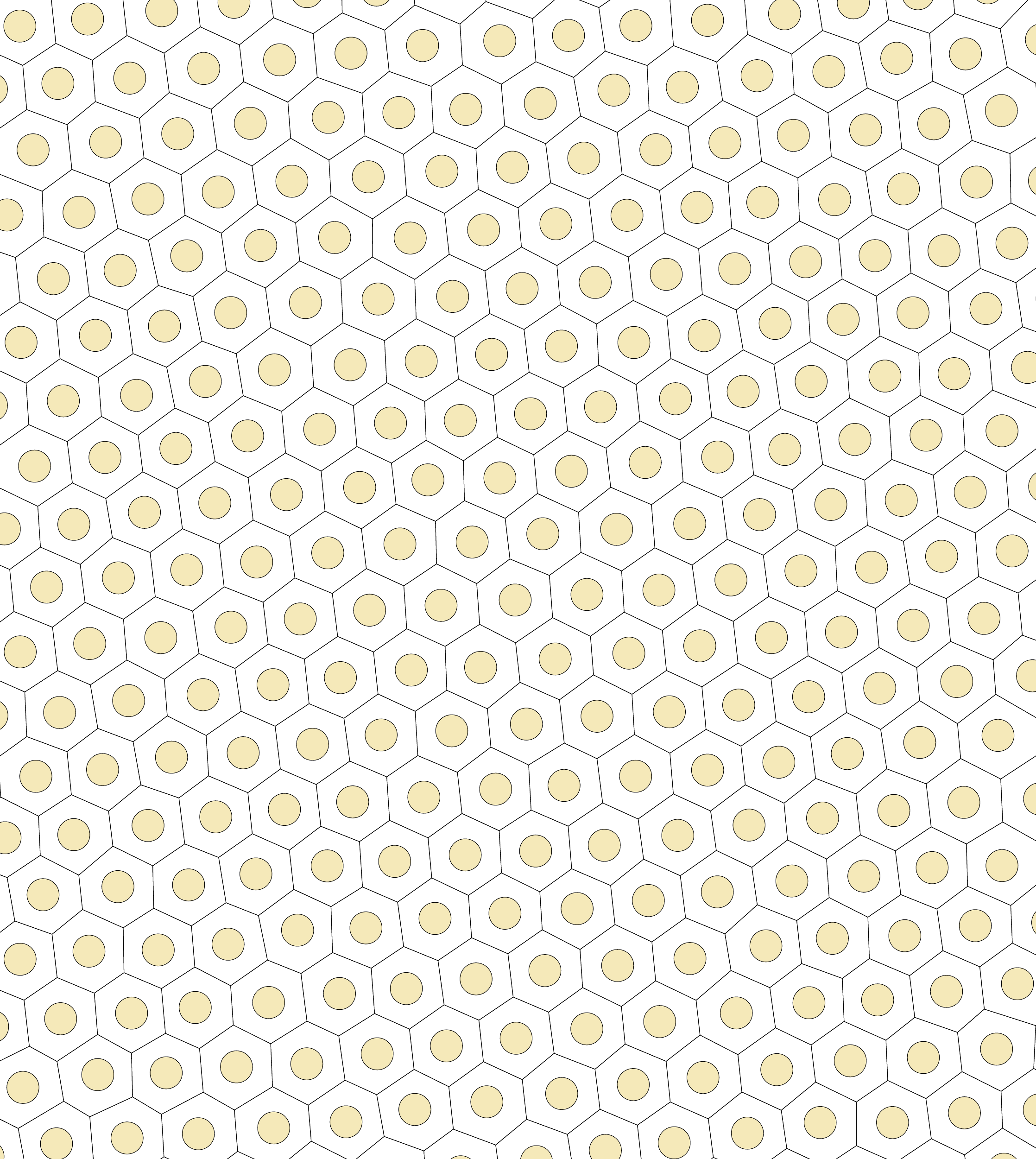}}&
\fbox{\includegraphics[trim={95mm 360mm 275mm 10mm},clip,height=0.42\columnwidth]{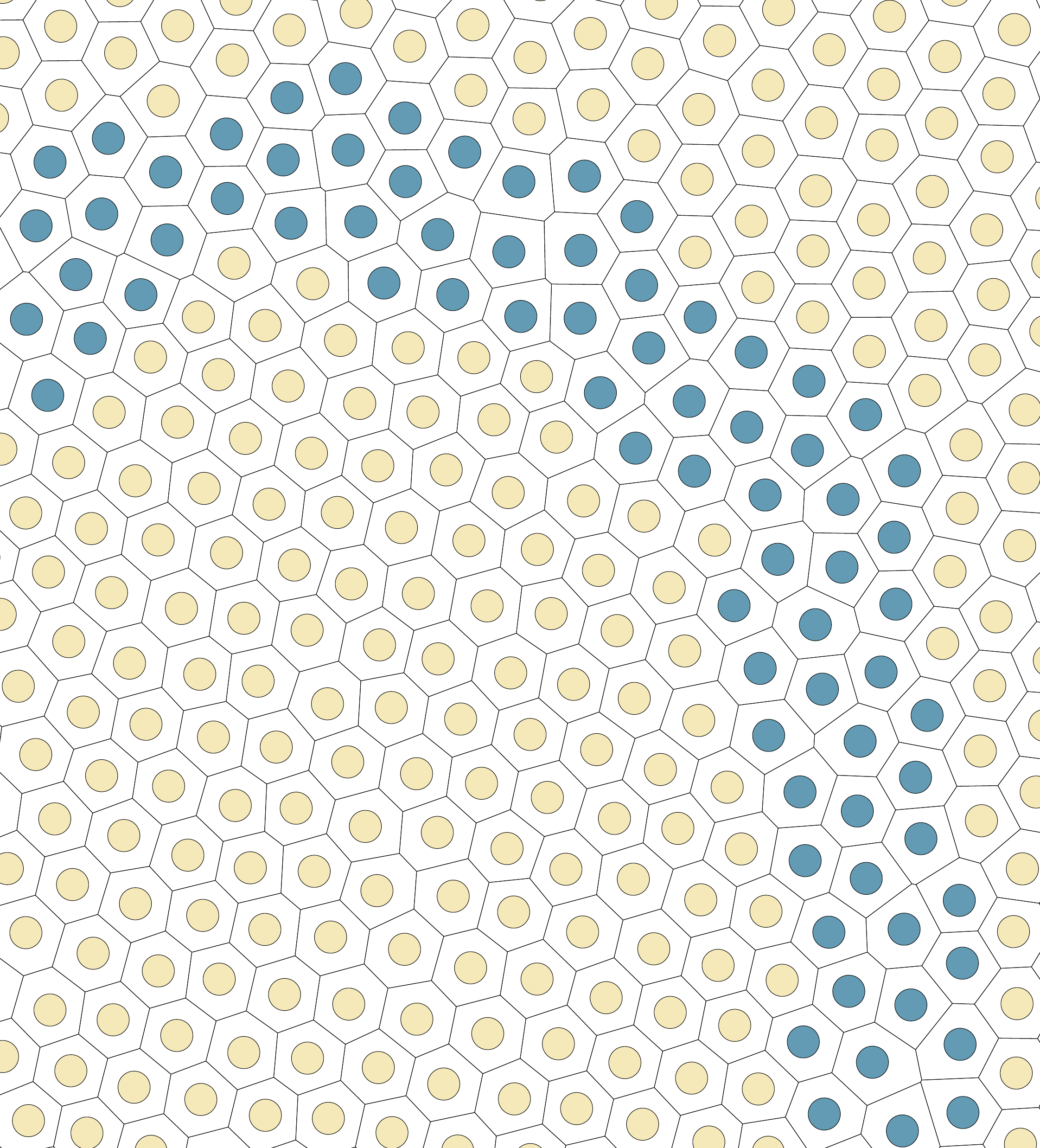}}&
\fbox{\includegraphics[trim={15mm 130mm 445mm 330mm},clip,height=0.42\columnwidth]{system1}}&
\fbox{\includegraphics[trim={220mm 190mm 160mm 190mm},clip,height=0.42\columnwidth]{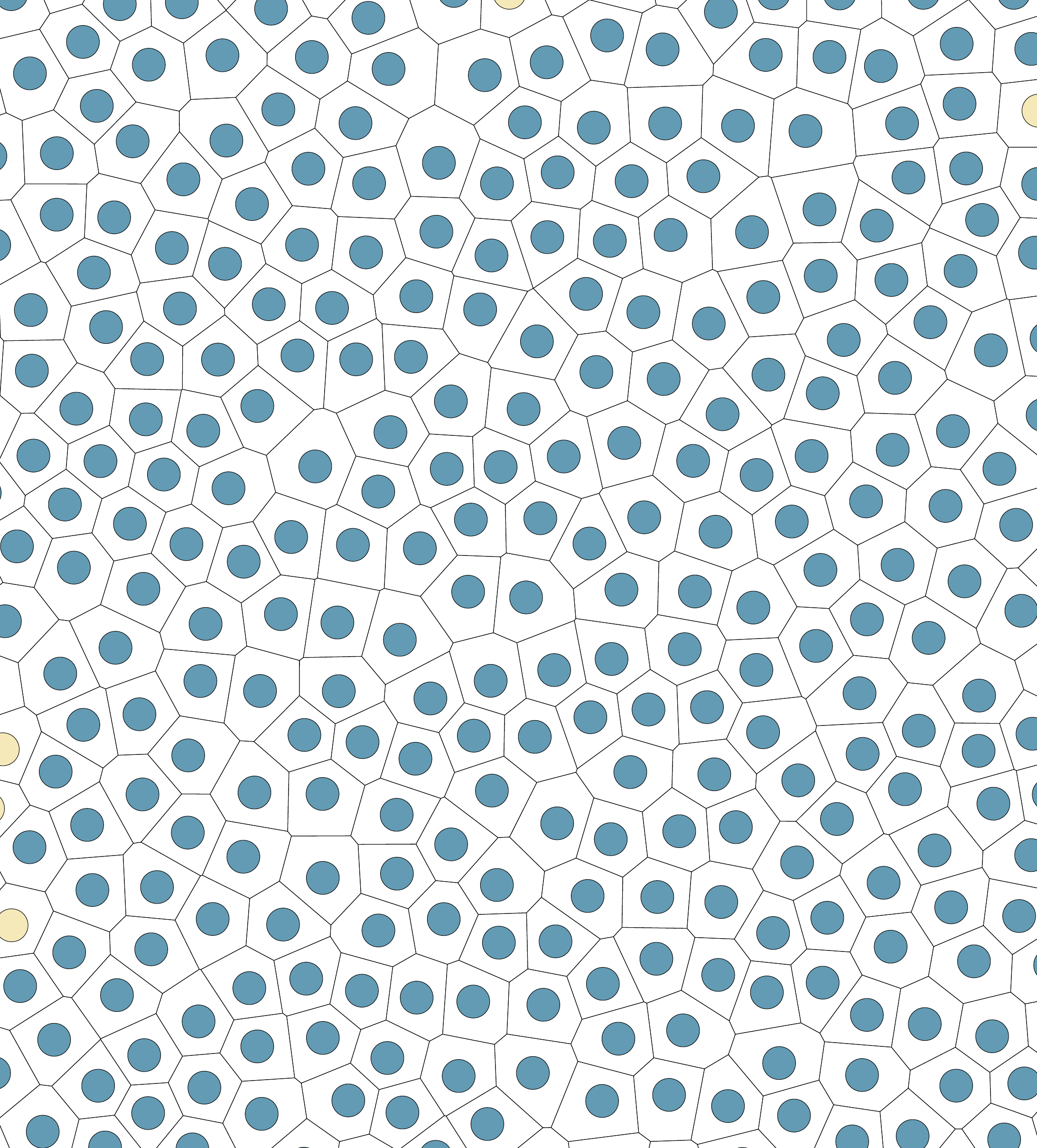}}&
\fbox{\includegraphics[trim={120mm 120mm 120mm 120mm},clip,height=0.42\columnwidth]{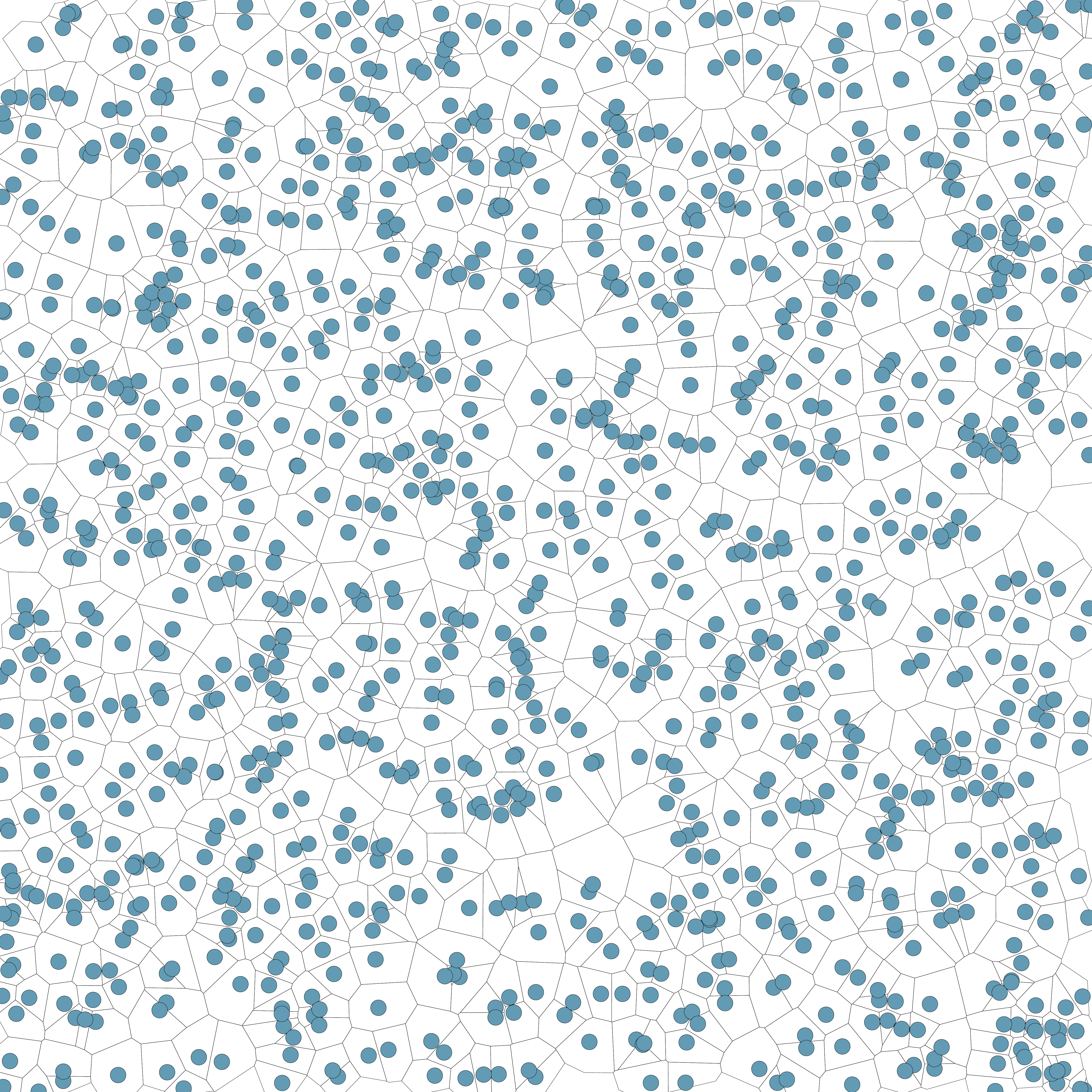}}\\
(a)&(b)&(c)&(d)&(e)
\end{tabular}
\caption{(a) Defect-free crystal, (b) two crystals meeting along a grain boundary, (c) crystal-liquid interface, (d) liquid, and (e) an ideal gas. Particles whose Voronoi cells and neighbors are all hexagons are colored yellow; all others are colored blue.
\label{fig:5systemsColor}}
\end{center}
\end{figure*}

There are computational methods for eliminating these discrepancies. For example,
the Voronoi cells can be computed using an exact arithmetic system (e.g., GMP~\cite{gmp_website}) that eliminates rounding error, albeit for
substantially higher computational cost. However, in many cases this problem
has limited effect on the results of interest. Cell-based computations such as
surface area and volume are not sensitive to small topological changes.
Furthermore, even though these numerical errors may introduce spurious faces
and edges, it is still possible to develop a topological framework in which
such errors are anticipated and accounted for (see Secs.~\ref{sec:topo} and \ref{sec:stability}).

Alternatively, algorithms exist that can compute the full Voronoi tessellation
as a single mesh, i.e., returning a full graph of blue edges as shown in
Fig.~\ref{fig:tessellations}. These approaches require building a global data
structure for the mesh, and thus are less amenable to parallelization when
compared to the cell-based perspective. When the Voronoi cells are
extracted from this mesh they will have internally consistent topology. Some
popular methods for computing the full Voronoi tessellation include the Fortune
sweeping algorithm\cite{fortune86,fortune87} and the incremental approach
whereby the mesh is continually updated as new particles are
added.\cite{green78,lee80} Another method is to use lift-up mapping,
projecting a particle $s\in \R^n$ to a paraboloidal surface $(s, \|s\|^2 ) \in
\R^{n+1}$. The hyperplanes tangential to the surface form facets  exactly
matching the Voronoi tessellation when projected back to $\R^n$. This method can be
efficiently implemented in arbitrary dimensions using the quickhull
algorithm.\cite{barber96}

A variety of software packages are available for computing the Voronoi
tessellation. The Qhull library\cite{qhull_website} computes Voronoi cells
using the quickhull algorithm, and forms the basis of Voronoi computations in
MATLAB. The Computational Geometry Algorithms Library
(CGAL)\cite{fabri2009cgal} has routines for calculating a variety of
geometrical constructions, including the Voronoi tessellation. Voro++ is a
software library for performing cell-based computations, and has both a
command-line utility and a C++ application programming interface for
calculating statistics of Voronoi cells.\cite{rycroft09a,rycroft09c} Voro++ is
incorporated into the OVITO software package for visualizing and analyzing
particle systems,\cite{stukowski10} and is integrated into the Large
Atomic/Molecular Massively Parallel Simulator (LAMMPS) for particle
simulation.\cite{lammps_website,plimpton1995fast,vollmayr2020introduction}
Several parallel implementations are also available. Starinshak et
al.\cite{starinshak2014new}\@ use a parallel divide-and-conquer approach, where
each processor computes Voronoi cells in a subdomain, after which they are
joined together. Other examples include PARAVT,\cite{gonzalez16} and a GPU
implementation.\cite{ray2018meshless}

\section{The shapes of Voronoi cells}

As the Voronoi cell of a particle is constructed using information about neighboring particles, its geometric and topological features in turn reflect important structural information about spatial arrangements of particles, both locally and globally.  We consider several examples.

\subsection{Geometric aspects}

In continuum mechanics, density is modeled as a property that varies continuously in space.
This viewpoint is difficult to justify when modeling systems at the atomic scale, because mass in such systems is localized at discrete points.
A reasonable question thus arises as to how density should be defined and analyzed, and the volume of a Voronoi cell is often used for this purpose.
In finite systems, the average area or volume of all Voronoi cells is equal to one divided by the average particle density.
The variation of particle density throughout a system can also be captured through the geometric properties of Voronoi cells.
Voronoi cells with small areas and volumes indicate particles in densely packed regions; conversely, those with large areas or volumes indicate particles whose neighbors are located far away. Section~\ref{sec:granular} discusses an example of using Voronoi volumes to analyze a granular material simulation.

The variation in the areas and volumes of Voronoi cells can also be indicative of structural abnormalities.  In low-temperature crystalline systems, for example, the Voronoi cells of bulk crystal particles are nearly identical.  Therefore, Voronoi cells with significantly larger or smaller areas and volumes can be indicative of localized defects such as vacancies and interstitials, as well as dislocations and grain boundaries.\cite{ashby1978structure}

In addition to geometric features of individual Voronoi cells, consideration of their distributions can also reflect information about global order.  In perfect crystals, for example, all Voronoi cells have identical volumes and surface areas.  At low temperatures, these distributions are narrowly concentrated around their mean values.  As a system is heated and becomes increasingly disordered, these distributions become broader.  These differences can be observed in Fig.~\ref{fig:5systemsColor}, which illustrates both ordered and disordered systems.
Although the Voronoi cell areas of particles in the crystals are nearly identical, there is considerable variance in those belonging to disordered regions.

\subsection{Topological aspects}
\label{sec:topo}

Topology is the mathematical study of connectivity and shape,\cite{willard2012general} and a thorough analysis of the topologies of individual Voronoi cells, as well as their distributions over entire systems, can provide much insight about local arrangements and global ordering.
Unlike geometric properties such as area and volume, topological ones such as the number of edges or faces of a Voronoi cell, are typically discrete, even non-numeric.
Indeed, the manner in which faces of a three-dimensional Voronoi cell are arranged relative to one another is captured through the isomorphism class of the cell's edge graph; this classification is typically represented in the language of graphs instead of numbers.\cite{2012lazar}

In solid-state physics, the unique Voronoi cell of a crystal lattice is often referred to as the \textit{Wigner--Seitz cell} of the system.\cite{wigner1933constitution,ashcroft1976solid}
In some two-dimensional crystals, for example, every Voronoi cell is a hexagon.  Particles whose Voronoi cells have more than or fewer than six edges can be identified as belonging to defects.  For example, Figs.~\ref{fig:5systemsColor}(a)--(c) illustrate crystalline systems in which most Voronoi cells have six edges.  In contrast, Voronoi cells of many particles lying along the grain boundary in Fig.~\ref{fig:5systemsColor}(b) have more or fewer edges.

In contrast to crystals, nominally disordered systems such as liquids and gases exhibit a substantially wilder set of particle arrangements, reflected in a larger variation of their topological features.  Instead of a single polyhedral-shaped tile centered at each lattice point (i.e., the Wigner--Seitz cell), disordered systems are tiled by a large assortment of Voronoi polyhedra.  All simple polyhedra, for example, appear with some frequency as Voronoi cells in the ideal gas,\cite{lazar2013statistical} providing an endless pool of Voronoi cell types for classifying particle arrangements.

The topology of a Voronoi cell in two dimensions can be characterized by its number of edges.  This classification can be refined by further considering the number of edges of the Voronoi cell's neighbors; such a classification is adopted in Fig.~\ref{fig:5systemsColor}.  The topology of a Voronoi cell in three dimensions is more delicate, due to the added complexity of describing the manner in which faces of a polyhedra are arranged.  This task can be done using an algorithm of Weinberg \cite{1966weinberg} originally designed to determine whether two planar graphs are isomorphic.  Because edge graphs of three-dimensional Voronoi cells are always planar,\cite{steinitz1922polyeder} Weinberg's graph-tracing algorithm provides a systematic description of Voronoi cell topologies.\cite{2012lazar,lazar2015topological}  This classification of local structure avoids some inherent limitations of more conventional methods.\cite{lazar2015topological,landweber2016fiber}  Furthermore, it facilitates the a priori enumeration of Voronoi topologies associated to particle structures.\cite{lazar2015topological}

The VoroTop \cite{vorotop} program is a command-line utility for calculating Voronoi topologies of three-dimensional particle systems.  The program uses the Voro++ software library\cite{rycroft09a} to compute the individual Voronoi cells, and then performs additional topological analysis.  By considering special families of Voronoi topologies associated with given crystalline systems, VoroTop facilitates the automated identification of defects in crystalline systems, as well as automated statistical characterization of disordered systems.

\subsection{Stability}
\label{sec:stability}

If properties of Voronoi cells are to be useful for classifying local structure, then it is important to understand the manner in which those properties can change under small perturbations of particle positions.
This issue is important to consider because measurement error is inevitable when working with experimental data.  Even numerical codes, in which particle coordinates are known precisely, are subject to rounding and approximation error.

The volume of a Voronoi cell is a continuous function of particle positions, and small perturbations of the particle coordinates thus result in small changes in cell volumes.  Likewise, most geometric features of Voronoi cells, such as perimeters and surface areas, are continuous functions of particle positions.\cite{reem2011geometric}  When geometric quantities are used for classification purposes, thresholds for the range of possible values must be chosen.  Even small errors in the measurements of particle positions can result in classification errors.

\begin{figure}
\begin{center}
\includegraphics[width=0.75\columnwidth]{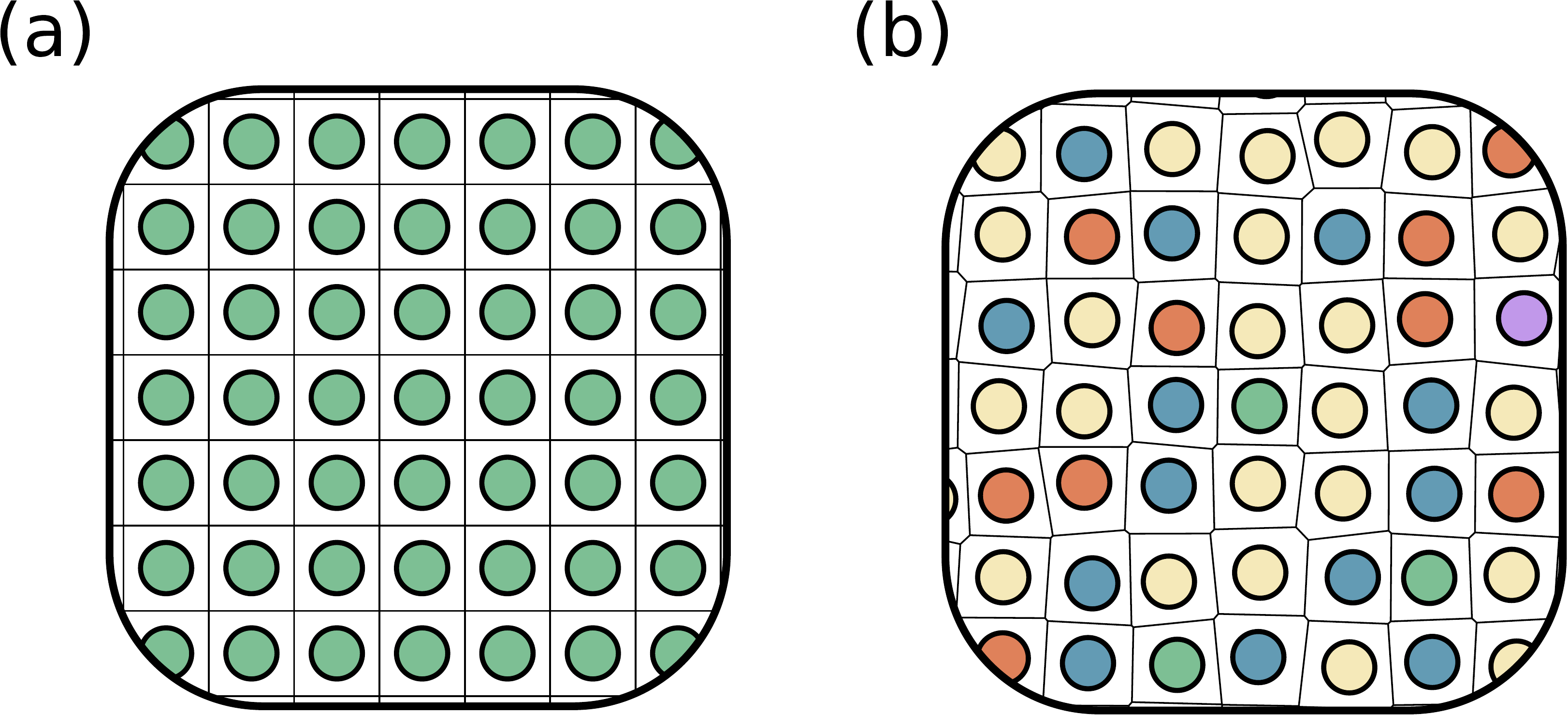}
\caption{Particles and their Voronoi cells in a (a) perfect and (b) perturbed square lattice.  Particles are colored according to the number of edges of their Voronoi cells.\label{fig:perturbedlattice}}
\end{center}
\end{figure}

Unlike geometric properties of Voronoi cells, topological ones can change abruptly under small perturbations of particle coordinates.\cite{weller1997stability,goberna2012stability}  For example, the number of edges or faces of a Voronoi cell can change under arbitrarily small perturbations of particle positions. Figure~\ref{fig:perturbedlattice} illustrates a perfect and perturbed square lattice in which particles are colored according to the number of edges of their Voronoi cells.  Small perturbations of the particles, which can be seen as representative of thermal vibrations or measurement errors, result in Voronoi cells that are topologically distinct from those in the unperturbed case.\cite{leipold2016statistical,lazar2020voronoi}  This instability arises from symmetries of the lattice which result in points equidistant to four neighboring particles. Small perturbations break this symmetry, resulting in topological changes.
\setlength{\fboxsep}{0pt}

\begin{figure}
\begin{center}
\includegraphics[width=0.57\columnwidth]{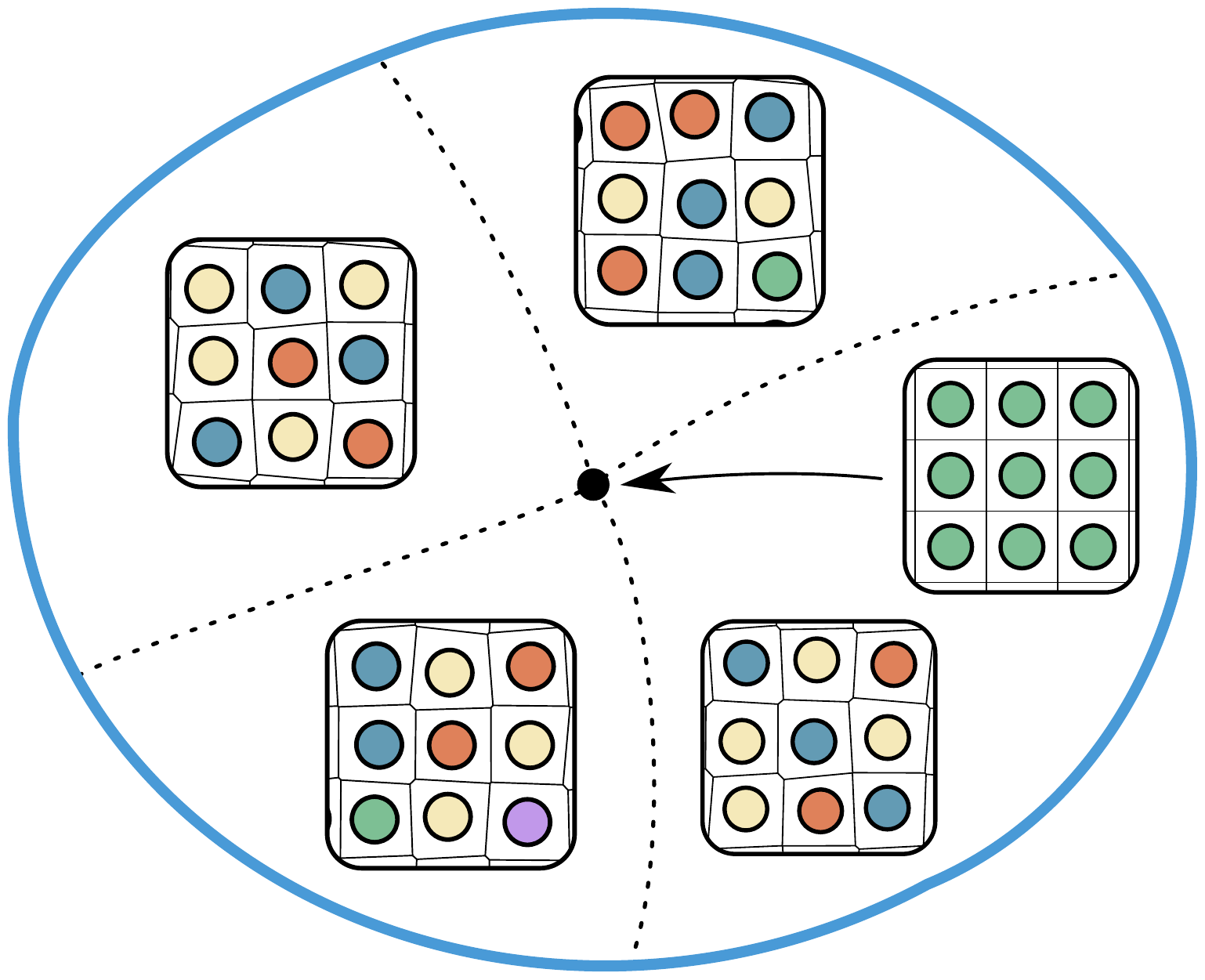}
\caption{Voronoi topology segments the configuration space of possible arrangements of particles into regions representing particles with a fixed kind of local structure.  Particles in unstable crystals correspond to points located at the boundary of several such regions.  Small perturbations of particle positions result in different Voronoi cell topologies.\vspace{-3.65mm} \label{fig:topspace_schematic}}
\end{center}
\end{figure}

This instability can be understood by considering an abstract {\it configuration space} of all possible local arrangements of particles,\cite{lazar2015topological} schematically illustrated in Fig.~\ref{fig:topspace_schematic}. Voronoi topology provides a natural means of segmenting this space into regions representing particles with a given local structure.  Particles in unstable crystals correspond to points located on the boundary of several such regions, and so perturbations of particle positions, corresponding to small perturbations in this space, typically result in different Voronoi cell topologies.    
The set of all topologies that can result from small perturbations can then be associated with a given structure type, allowing for a simple method for dealing with experimental error and thermal vibrations.\cite{lazar2015topological}  Combinatorial factors limit the number of types that can result from small perturbations, corresponding to the number of regions meeting at the representative point in the configuration space.  
In many cases, a complete set of types that can result from small perturbations can be determined analytically or numerically.\cite{lazar2015topological,lazar2018}

Not all lattices and crystals are unstable in this sense.  In two dimensions, for example, the hexagonal lattice, whose unique Voronoi cell is a regular hexagon, is stable under small perturbations; cf., Fig.~\ref{fig:5systemsColor}(a).  This arrangement of particles corresponds to a point in the interior of a region of configuration space with fixed Voronoi topology.  In three dimensions, the body-centered cubic lattice is stable, whereas the majority of common crystals, such as the simple cubic, face-centered cubic, hexagonal close packed, and diamond crystals, are not.\cite{lazar2015topological}

\section{Applications}

In recent decades, Voronoi tessellations and their generalizations have been used to analyze countless problems throughout the physical sciences, as well as in the life sciences, engineering, and pure mathematics.  
For example, Voronoi tessellations have found broad use in 
computational fluid dynamics,\cite{weatherill1992delaunay,springel2011hydrodynamic,bres2018large} 
secure data compression and transmission,\cite{ahuja1985image,rom1988image,du2006centroidal,melkemi2020voronoi}
machine learning,\cite{heath1993complexity,khoury2019adversarial,fukami2021global}
quantum error correction,\cite{harrington2004analysis,albert2020robust}
and a wide range of biological, chemical, and ecological problems.\cite{sussman2018anomalous,shahmoradi2016dissecting,votel2009equilibrium,stewart2010voronoi}
Even a brief survey of the dozens of such applications is well beyond the scope of this article.  Instead, we mention several examples of particular interest to physicists and detail how Voronoi tessellations are used in their analysis.

\subsection{Crystals and defects}

Many common materials such as diamonds, table salt, and most everyday metals, are crystalline in nature. Although the bulk of these materials are highly ordered, the underlying structure is punctuated by defects of various dimensions and kinds.\cite{ashcroft1976solid,kittel1976introduction} Material scientists wish to understand these defects, because they impact macroscopic material properties such as strength and ductility.\cite{callister2000fundamentals}  To study defects, computational material scientists often use supercomputers to simulate systems with millions, or even billions, of atoms.  However, without a way to automate the accurate identification of the defects, the massive amounts of data is not useful.  Automating the analysis of large atomistic data sets thus motivates the development of methods to identify and classify defects using local structural data.

\begin{figure}
\begin{center}
\begin{tabular}{ccc}
\fbox{\includegraphics[angle=90,trim={380mm 450mm 200mm 160mm},clip,width=0.315\columnwidth]{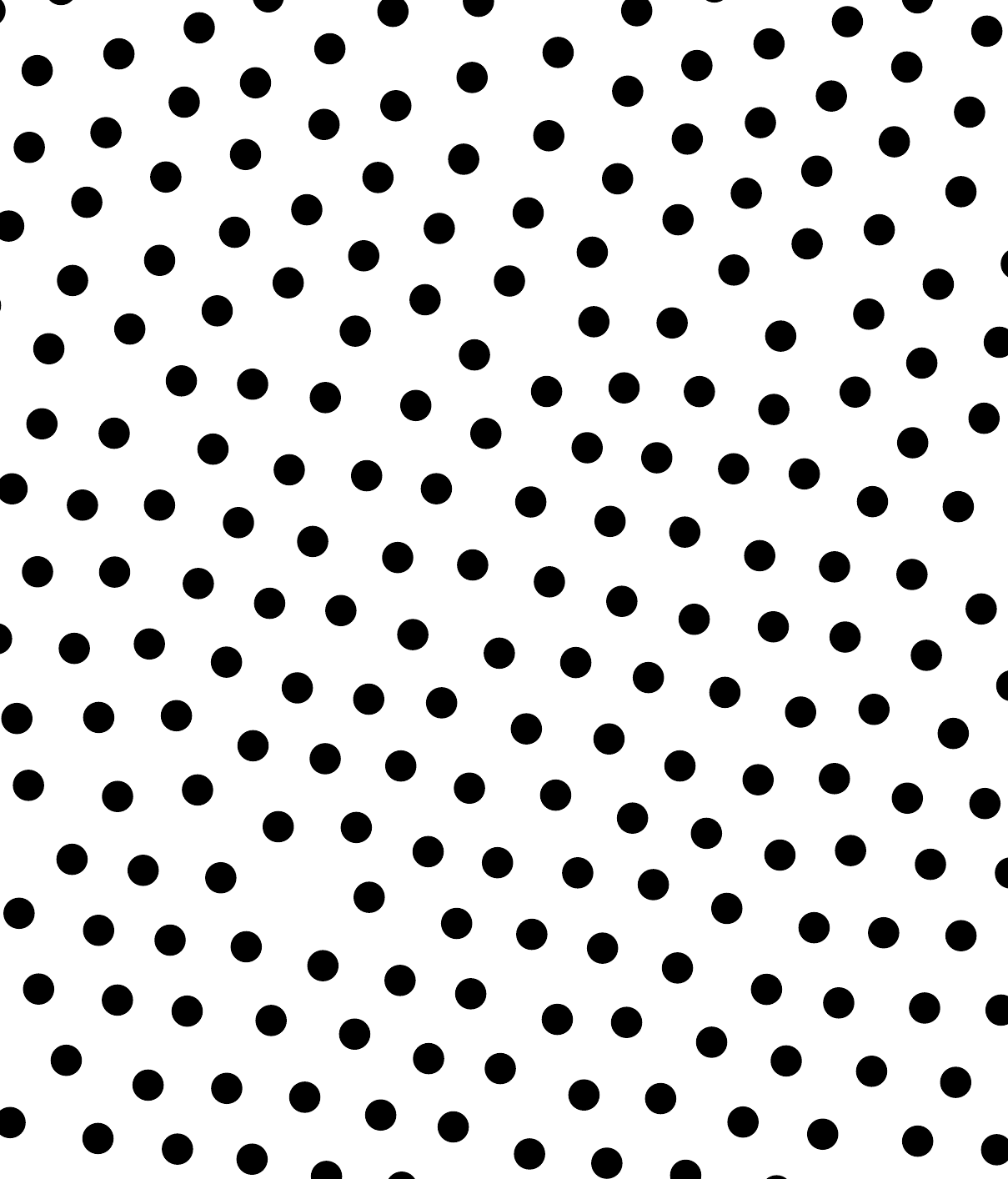}}& 
\fbox{\includegraphics[angle=90,trim={380mm 450mm 200mm 160mm},clip,width=0.315\columnwidth]{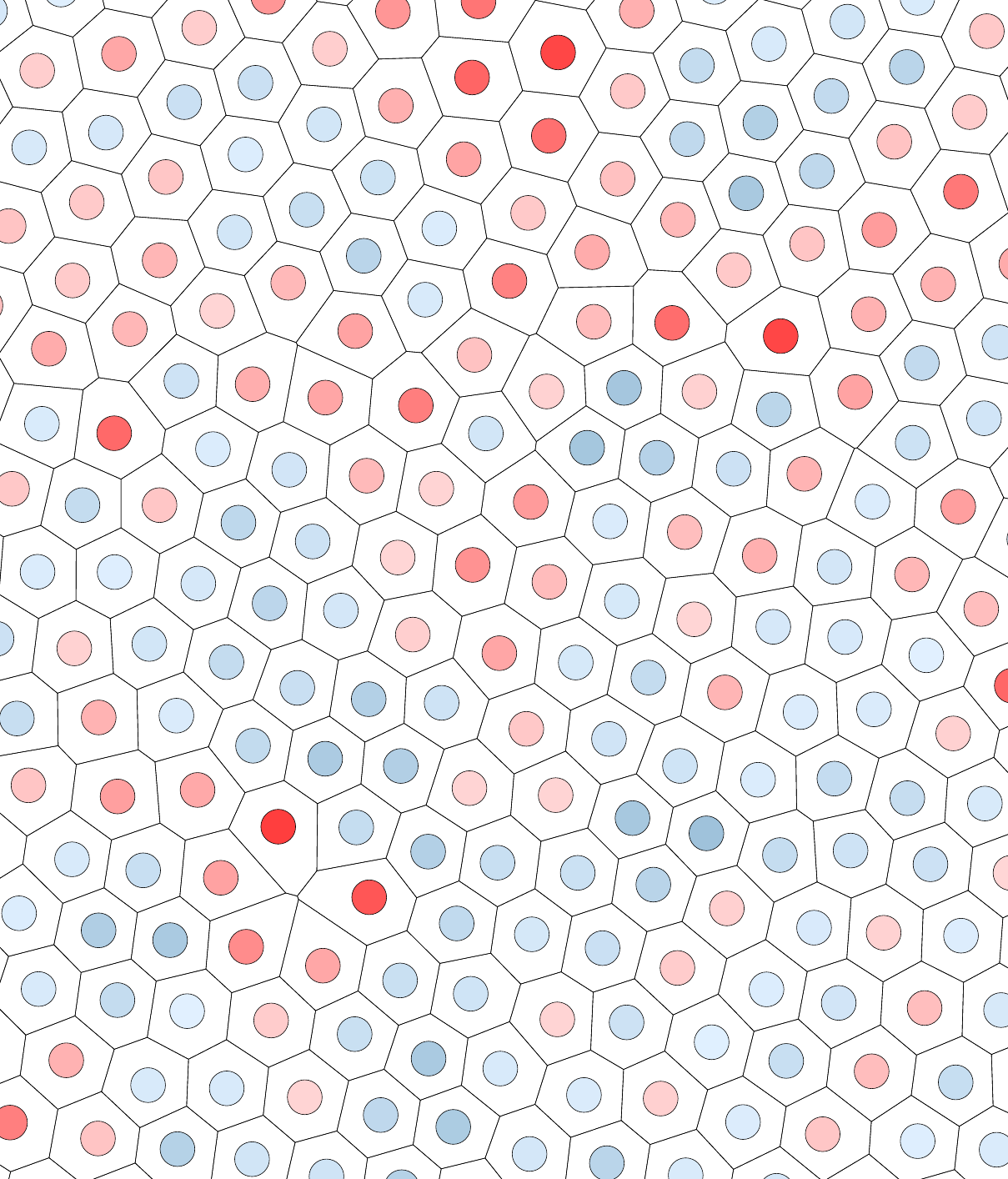}}& 
\fbox{\includegraphics[angle=90,trim={380mm 450mm 200mm 160mm},clip,width=0.315\columnwidth]{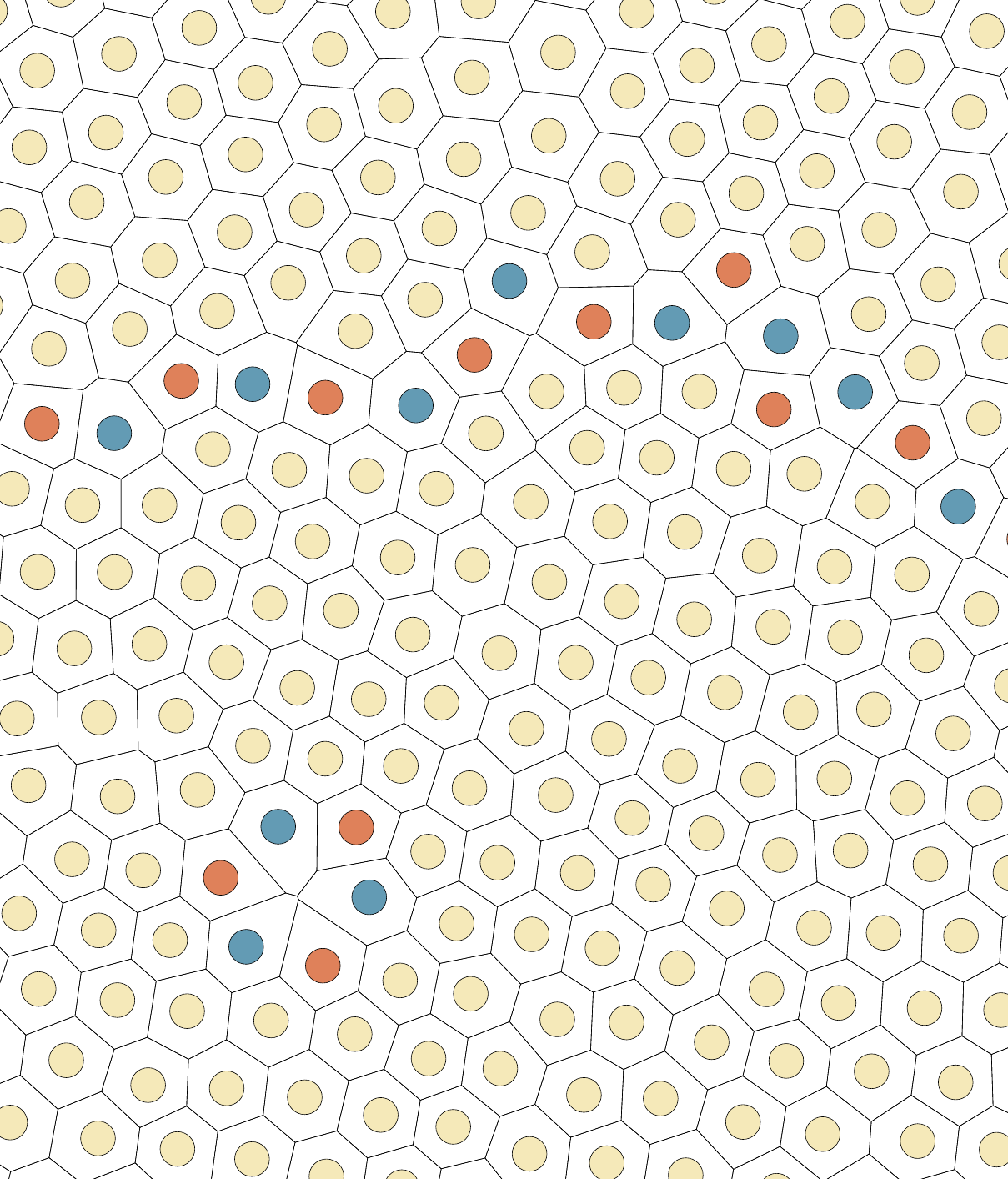}}\\
(a)&(b)&(c)
\end{tabular}
\caption{(a) Two adjacent crystals and a vacancy defect. (b) Particles are colored according to the areas of their Voronoi cells.  Those with larger areas are colored red, and those with smaller ones are colored blue. (c) Particles are colored according to the number of Voronoi cell edges.
\label{fig:bicrystal_with_vacancy}}
\end{center}
\end{figure}

Figure \ref{fig:bicrystal_with_vacancy} illustrates a two-dimensional system with two crystals, a boundary between them, and a localized vacancy.  Particles are colored according to their Voronoi cell areas and topologies in Figs.~\ref{fig:bicrystal_with_vacancy}(b) and \ref{fig:bicrystal_with_vacancy}(c), respectively.  Although the vacancy defect is associated with Voronoi cells with above-average areas, corresponding to a decrease in the local density, area changes resulting from thermal fluctuations of other particles in the system make it difficult to identify the defects by the analysis of Voronoi cell areas alone.  When colored according to Voronoi topologies, particles belonging to the two bulk crystals appear locally ordered, and their Voronoi cells all have six edges. Those adjacent to the grain boundary and vacancy have more than or fewer than six edges.  This observation suggests using the topology of Voronoi cells to identify, and possibly classify, defects in crystalline systems.

\setlength{\fboxsep}{0pt}
\setlength{\tabcolsep}{0.2em}

Although this example is two-dimensional, this approach is also effective in three dimensions, and enables the accurate identification of defects in crystalline systems at high temperatures, where conventional methods fail.\cite{lazar2015topological,vorotop,lazar2018}

\subsection{Disordered systems}

Although originally introduced to study crystalline systems, Voronoi cells also have a long and storied history in the study of liquids, glasses, and other nominally disordered systems.  In the early 1930's, John D.~Bernal, the great pioneer of X-ray diffraction and founder of protein crystallography, initiated a study of simple liquids.\cite{bernal1933theory, fowler1933note, bernal1937attempt}
Bernal suggested looking at the Voronoi cells of particles, and considering not only their number of faces but also the number of edges of each face.  This detailed analysis of Voronoi cells enabled him to describe the structure of liquids through a topological description of their constituent particles.

Despite lacking crystalline order, nominally disordered systems can still be distinguished from one another and analyzed through local structural features.  For example, liquid copper and an ideal gas are both disordered, and yet are structurally distinct.  One of the challenges in studying disordered systems is meaningfully describing and accurately identifying different kinds of structural order.

The distribution of Voronoi topologies can be effectively used for this purpose.\cite{montoro1993voronoi,cheng2013local,derlet2020correlated,han2020atomistic} In particular, a detailed list of the Voronoi topologies and their observed frequencies provides important structural insight into local structure character.  Icosahedral symmetry, studied in the context of liquids and glasses,\cite{jonsson1988icosahedral} is but a single type of local structure, and has shed light on the tension between local energy-minimizing and entropy-maximizing principles in determining how particles arrange in disordered systems. Although our understanding of the distribution of local Voronoi topologies in general systems is still incomplete, this data has been recently used to identify phase transitions in supercritical fluids.\cite{yoon2018topological,yoon2019topological,yoon2019topological2}

\subsection{Granular flows}
\label{sec:granular}

\begin{figure*}
  \begin{center}
    \includegraphics[width=\textwidth]{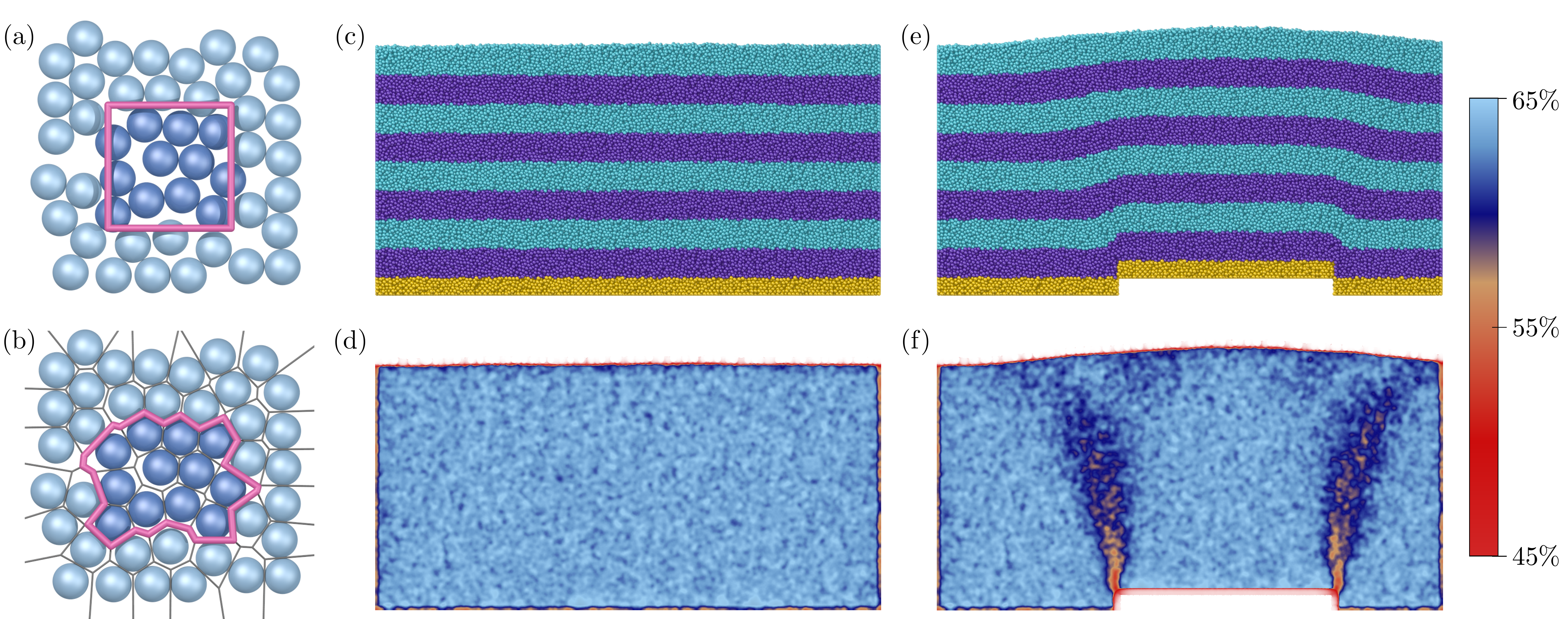}
  \end{center}\vspace{-0.5em}
  \caption{(a) A simple method to estimate local volume fraction is based on counting the particles in a small control volume (purple square). (b) By modifying the control volume to be given by the Voronoi cells of the particles, the local volume fraction can be calculated more accurately. (c) Snapshot of a granular discrete-element simulation using 140,000 particles of diameter $d$ poured into a container of width $140d$ and thickness $12d$. The yellow particles are frozen in place, and the dark blue and light blue particles are colored in layers of $8d$. (d) Local particle density calculated using Voronoi volumes; see color key on right. (e) Snapshot of particles after a section of the frozen particles has been slowly raised by $4.7d$. (f) The corresponding local particle density, showing two bands of lower density; see color key on right.\label{fig:gran}}
\end{figure*}

Voronoi tessellations have also been successful in characterizing the dynamics of granular flows. Despite being common in everyday experience, granular materials have surprisingly complex mechanical properties: in some situations they behave like solids and support stress, whereas in other situations they can flow like liquids.\cite{jaeger96} This complex behavior has prompted much study in the past several decades to fully understand the mechanics of granular media.\cite{jop06,henann13} A key quantity of interest is how closely packed the particles are.  If they are tightly packed, they will become jammed\cite{ohern03} and behave like a solid, but in order to flow, there must be some free volume for the particles to move past each other. The notion of free volume has been extensively studied for many amorphous materials\cite{turnbull61,cohen79} and has also been used as the basis for simple models of granular drainage.\cite{caram91,rycroft10}

Voronoi cells are useful for precisely characterizing the local volume fraction, i.e., what proportion of space is occupied by the particles. A simple approach is shown in Fig.~\ref{fig:gran}(a), where a small pink square of volume $V_s$ is drawn. Twelve particles with volume $V_p$ have centers inside the square. Thus the local volume fraction can be estimated as $12V_p/V_s$. However, this measurement is sensitive to small changes in the particle positions: if one particle's center moves outside the box, the volume fraction will become $11V_p/V_s$, which is a $8\%$ relative change. This change is problematic, because small fluctuations of the packing fraction of several percent can have a large effect on the mechanical properties. This calculation can be improved by computing the Voronoi cells for the particles. The pink square volume can be replaced by the total volume $V_v$ of the Voronoi cells [Fig.~\ref{fig:gran}(b)], resulting in a more faithful representation of the space associated with the particles, and a more accurate computation of the local volume fraction.

To illustrate this method, we performed a granular discrete-element simulation, where the position, velocity, and angular velocity of each particle are integrated using Newton's laws. We simulate 140,000 particles of diameter $d$ using LAMMPS,\cite{lammps_website,plimpton1995fast} where parameters in the inter-particle contact model are derived from Rycroft et al.,\cite{rycroft12c} with a Coulomb friction coefficient of 0.5\@. Particles are first poured into a container of width $140d$ and thickness $12d$, using periodic boundary conditions in the thickness direction. Figure~\ref{fig:gran}(c) shows a snapshot of the simulation, where the bottom layer of particles shown in yellow are frozen in place. The remaining particles are all physically identical but are colored in alternating bands of width $8d$. Figure~\ref{fig:gran}(d) shows the local packing fraction, achieving a value of $\approx 63\%$ which is typical of random close packing.\cite{torquato_book}

A section of the frozen particles are then slowly raised a distance of $4.7d$. Figure~\ref{fig:gran}(e) shows how the layers of particles are deformed, but it is difficult visually to see any change in the packing fraction. However, the Voronoi-based calculation demonstrates that two bands of lower packing fraction form in regions undergoing high shear, where particles must move apart slightly in order to move past one another. Such analyses can be used to probe the mechanics of granular media.\cite{rycroft09b}

Voronoi volumes have been used to probe a variety of aspects of granular media. Although Fig.~\ref{fig:gran}(b) demonstrates the computation of the  local packing fraction in a small region, the same principle can be used to define the local packing fraction at the level of each particle, by dividing the particle volume by its Voronoi cell volume. This approach has been used to examine local volume fluctuations~\cite{rycroft06a,puckett11,suikkanen14,gago16} and dynamic jamming fronts.\cite{waitukaitis13} The use of Voronoi cell faces to identify neighboring particles\cite{panaitescu14} and the examination of Voronoi cell anisotropy~\cite{guo14,rieser16} have also proven fruitful in the study of granular media and other amorphous systems.

\subsection{Astronomy} 

In the examples we have considered so far, the particles around which the Voronoi cells are constructed have represented atoms, or other small-scale objects.  Voronoi tessellations have also found applications when the particles under consideration are of astronomical scale.  In an attempt to understand the Universe and its origins, astronomers have spent recent decades imaging and cataloging the observable sky, carefully mapping out its stars, galaxies, and other celestial bodies.  Dozens of digital surveys such as the Two Micron All Sky Survey\cite{skrutskie2006two} and the Sloan Digital Sky Survey\cite{york2000sloan} 
have generated many petabytes and exabytes of data, from which scientists have culled novel insight into many secrets of the cosmos.

To fully exploit the breathtakingly large amount of recorded data, scientists have been eager to develop automated computational techniques to aid in its analysis.
Voronoi tessellations have consequently been used to study the large scale distribution of galaxies,\cite{vavilova2021voronoiB} as well as to identify individual astronomical objects such as
galaxy groups and clusters,\cite{ramella2001finding} haloes,\cite{neyrinck2005voboz} voids,\cite{neyrinck2008zobov} and filaments.\cite{gonzalez2010automated}
Each of these studies has provided insight into the early history of the universe, its evolution, and its large-scale structure.

\subsection{Numerical analysis}
\label{sub:num_an}

\begin{figure*}
  \begin{center}
    \includegraphics[width=\textwidth]{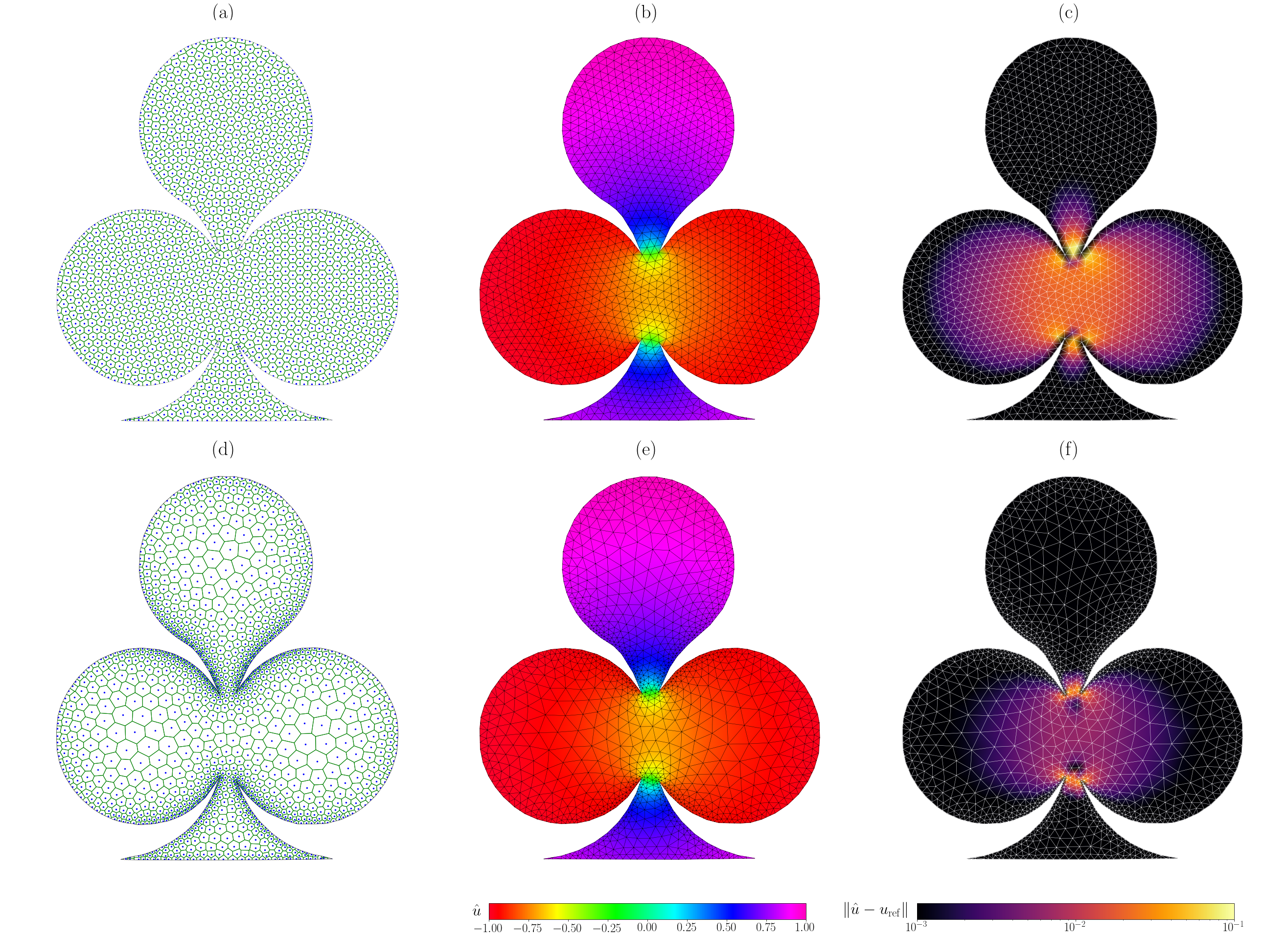}
  \end{center}\vspace{-1.2em}
  \caption{(a) The Voronoi diagram associated with a uniform triangular mesh of the club shape domain $\Omega$, using $n=1468$ points. (b) The corresponding Delaunay triangular mesh for the uniform mesh. The numerical solution $\hat{u}$ of the Laplace equation $\Delta u = 0$ is shown on the uniform mesh. (c) The absolute errors $\| \hat u-u_\text{ref}\|$ of the numerical solution for the uniform mesh. (d) The Voronoi diagram for the adaptive mesh, using $n=1468$ points. (e) The corresponding Delaunay triangulation and the numerical solution of the Laplace equation on the mesh. (f) The absolute errors of the numerical solution for the adaptive mesh.\label{fig:final_plot}}
\end{figure*}

One important problem that arises in modeling countless physical systems is the numerical solution of partial differential equations on domains with nontrivial shapes. A necessary first step in solving many such problems is the discretization of the domain, and Voronoi tessellations are commonly used for this purpose.\cite{du2002gridgen,du2003sizingdensity,ju2006adaptiveFEM,ringler2011climateCVD} Here we illustrate how to solve one such problem using both a uniform mesh and an adaptive one, both constructed using Voronoi tessellations.

As shown in Fig.~\ref{fig:final_plot}(a), our computational domain $\Omega$ is the club shape from a standard playing card deck, which we chose due to its sharp corners and concave creases. We use the finite element method\cite{claes_fem} to solve the two-dimensional Laplace equation
\begin{equation}
\label{eqn:poissons_equation}
\Delta u(x,y)=0,
\end{equation}
subject to a Dirichlet boundary condition on the boundary $\p \Omega$. To test the accuracy of different meshes, we use the method of manufactured solutions \cite{salari2000manufacturedsolution,roache2002manufacturedsolution} to build a reference analytical solution of the problem. Because the imaginary part of an analytic function is harmonic, and is thus a solution to Laplace's equation, we construct an exact reference solution as
\begin{equation}
  u_\text{ref}(x,y) = b+c\operatorname{Im} \left( \sum_{k=1}^4 \alpha_k [(x+yi) - (x_k+y_ki)]^{1/6} \right),
\end{equation}
where $(x_k,y_k)$ are the four corners of the concave creases of the club shape, $\alpha_k$ are arbitrary constants, and we choose the branch cuts of the fractional powers to not intersect with $\Omega$. The constants $b$ and $c$ are chosen to normalize $u_\text{ref}$ to have maximum value $1$ and minimum value $-1$ on $\Omega$. The solution $u_\text{ref}$ is smooth in $\Omega$, and varies most rapidly near the four concave creases.

We aim to solve Laplace's equation where the Dirichlet condition is set to $u_\text{ref}$ on the boundary $\partial \Omega$. Because the solution to the Laplace equation is unique in this case, it must equal $u_\text{ref}$ everywhere on $\Omega$. Thus we can compare our numerical solution $\hat{u}(x,y)$ to $u_\text{ref}$ to test the accuracy of the numerical method. We use a piecewise linear triangular mesh.\cite{claes_fem} To generate such a mesh, we first create a Voronoi tessellation and then use it to construct the dual Delaunay triangulation by connecting points that share a Voronoi edge.\cite{loera} The Delaunay triangulation has the property that for any triangle, no other points are in the circumcircle of the triangle. Delaunay triangulations are attractive for use with finite element methods, because the definition disfavors thin triangles that have small internal angles and lead to large numerical errors.

\begin{figure}[b]
\begin{center}
\includegraphics[width=\columnwidth]{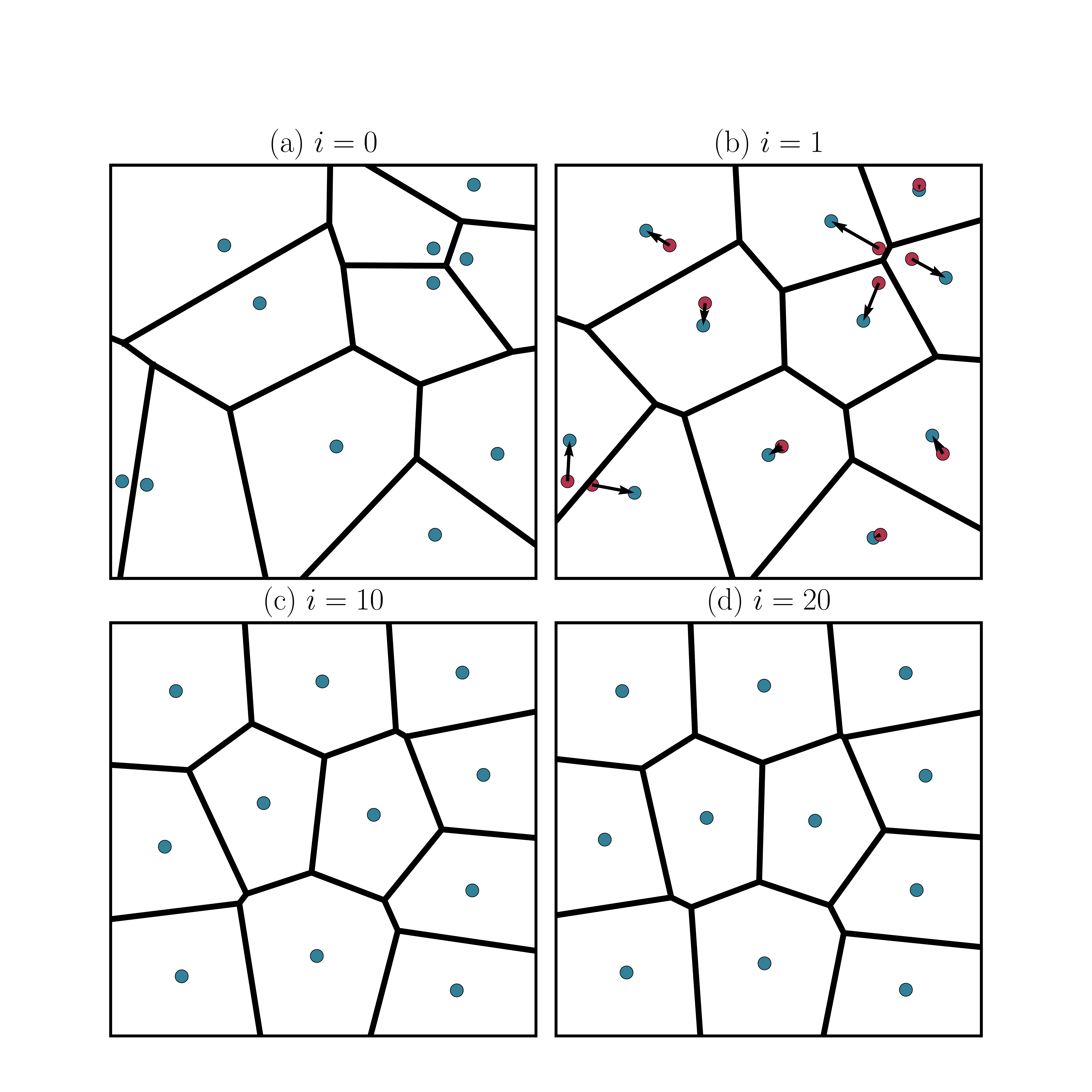}\vspace{-0.5em}
\caption{A simple illustration of Lloyd's algorithm. (a) The initial random points and the associated Voronoi diagram. (b) After one iteration, the points moved to the centroids of their Voronoi cells, as indicated by the directional arrows, where the red points were the old point locations, and the blue points are the current point locations. The blue points form a new Voronoi diagram. (c) The points and their associated Voronoi diagram after $10$ iterations. (d) The points and their associated Voronoi diagram after $20$ iterations. Compared to (a), the points distribute more uniformly in the domain, and their Voronoi cells are of similar sizes. \label{lloyd_illustration}}
\end{center}
\end{figure}

We therefore need to compute a Voronoi tessellation of $\Omega$ using evenly distributed points.  To do this, we use a \textit{centroidal Voronoi diagram},\cite{du1999centroidal} where each point coincides with the weighted center of mass, or centroid, of its Voronoi cell. A centroidal Voronoi diagram can be computed via Lloyd's algorithm.\cite{lloyd1982,du1999centroidal} For an initial arbitrary set of points, Lloyd's algorithm iteratively movies the points to transform the corresponding Voronoi diagram into a centroidal Voronoi diagram. The algorithm is as follows:
\begin{enumerate}
  \item Choose $n$ initial points and compute their associated Voronoi diagram.
  \item Compute the centroid for each of the Voronoi cells.
  \item Move each point to its Voronoi cell's weighted centroid location.
  \item Repeat steps (2) and (3) until a stopping criterion is satisfied, such as a maximum number of iterations or a minimum distance of point movement.
\end{enumerate}
Figure \ref{lloyd_illustration} illustrates how Lloyd's algorithm works.  As seen in Fig.~\ref{lloyd_illustration}(a), we first choose $11$ random points in the unit square and compute their corresponding Voronoi diagram. We then implement Lloyd's algorithm for $20$ iterations. After $20$ iterations, as seen in Fig.~\ref{lloyd_illustration}(d), we obtain a Voronoi diagram where the points are uniformly distributed in the domain, and their Voronoi cells are of similar sizes.

Figure~\ref{fig:final_plot}(a) shows a centroidal Voronoi diagram of $\Omega$, using $n=1468$ points. From here, the corresponding Delaunay triangulation can be computed, and the finite element method can be implemented.  The numerical solution $\hat{u}$ is shown in Fig.~\ref{fig:final_plot}(b). Figure~\ref{fig:final_plot}(c) shows the absolute errors, $|\hat{u}(x,y)-u_{\text{ref}}(x,y)|$, for the numerical solution, indicating that the largest errors are near the four concave creases. This is expected because the solution varies most rapidly there.

To improve the accuracy, we can modify our mesh to use adaptive resolution. To do so, we introduce a spatially-varying density field $\rho(x,y)$ in the computation of the Voronoi cell centroids. Specifically, the weighted centroid $\mathbf{c}$ of a Voronoi cell $V$ is
\begin{equation}
\label{eqn:voro_centroid}
\mathbf{c}=\frac{\int_V \mathbf{x} \rho(\mathbf{x}) d\mathbf{x}}{\int_V \rho(\mathbf{x}) d\mathbf{x}},
\end{equation}
where $\mathbf{x}=(x,y) \in \mathbb{R}^2$. Equation \eqref{eqn:voro_centroid} requires computing integrals over $V$. We do this by dividing the Voronoi cell into a set of triangles and using a quadrature rule on each of them.\cite{heath,jane2010CVTconstraints}

The density field can be set manually, but here we use an automatic procedure that forces the triangles in the interior of the domain to be large, the triangles near the boundary to be small, while ensuring a smooth gradation in mesh size.\cite{alliez2005sizingfield,jane2010CVTconstraints, du2003sizingdensity, du2006sizingdensity} Figure~\ref{fig:final_plot}(d) shows the resulting Voronoi tessellation using the same number of $n=1468$ points.  The points of the resulting CVD distribute more densely around the corners and the concave creases of the shape, and less densely in the interior of the three lobes.
Figure~\ref{fig:final_plot}(e) shows the corresponding Delaunay triangulation and the finite-element solution.
Because the mesh resolution is increased in the regions where $u_\text{ref}$ varies most rapidly, the numerical
errors are reduced when compared to the uniform mesh, as shown in Fig.~\ref{fig:final_plot}(f).

We can also compare the errors of the two cases by defining an $L^1$ error norm,
\begin{equation}
\label{eqn:error_integral_measure}
\|e\|_1= \frac{1}{A} \int_\Omega |\hat{u}(x,y)-u_{\text{ref}}(x,y)| \,dx\,dy,
\end{equation}
where $A$ is the area of $\Omega$. The integral in Eq.~\eqref{eqn:error_integral_measure} is computed numerically via a quadrature rule on the Delaunay triangles. The resulting $\|e\|_1$ for the adaptive mesh is $0.001459$, and $0.004764$ for the uniform mesh. Hence, by using an adaptive mesh instead of a uniform one, we can improve the accuracy of the numerical solution by a factor of three for the same amount of computational work.

\section{Conclusions}

Voronoi tessellations provide a natural framework for describing and analyzing arrangements of  particles on a wide range of scales. Careful analysis of geometric and topological features of their constituent polyhedra-shaped cells can provide insight not readily apparent from other perspectives.

Despite being introduced over a century ago as an abstract mathematical concept,  Voronoi tessellation continues to find new applications in many contemporary research problems. In the last decade there has been much interest in data-driven approaches in science, such as the analysis of large databases,\cite{pablo19} and the use of new machine learning methods.\cite{zdeborova17} These approaches often rely on the construction of simple descriptors of a system, and Voronoi cell features have proven highly useful, such as for screening high-throughput databases~\cite{willems12,lin12} and for predicting properties of crystalline structures.\cite{ward17}

\setlength{\unitlength}{0.001\textwidth}
\begin{figure}
  \begin{center}
    \begin{picture}(474,144)
      \put(-2,0){\includegraphics[width=474\unitlength]{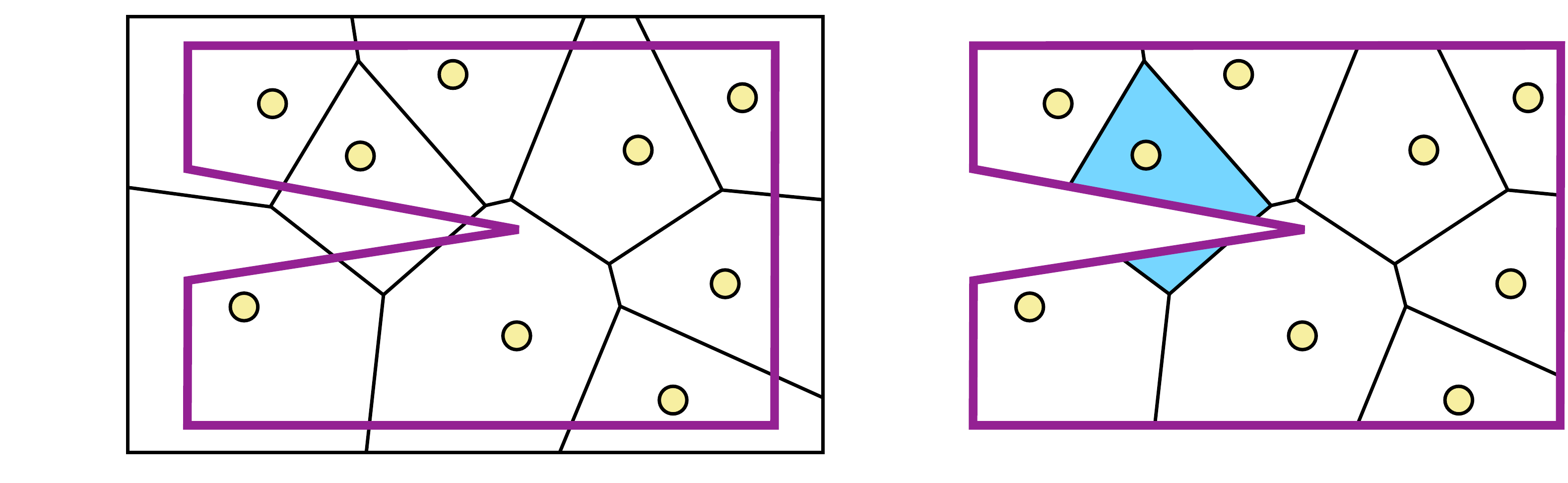}}
      \put(7,120){(a)}
      \put(263,120){(b)}
    \end{picture}
  \end{center}\vspace{-4mm}
  \caption{(a) For the nine points shown by the yellow circles, the Voronoi tessellation in a rectangular domain is shown in black. The outline of a domain $U$ is shown in purple. (b) If the Voronoi tessellation is restricted to $U$, then one of the Voronoi cells, shown in blue, is split into two disconnected components.\label{fig:voro_bdry}}
\end{figure}

\section{Suggested problems}

We suggest several problems suitable for motivated students, perhaps as part of a course in computational physics.  Those marked with a $^{\star}$ might form the basis of a semester-long research project.

\begin{enumerate}[label={(\arabic*)}]
\item Accurately measuring particle positions in experimental systems can be difficult. Drawing quantitative conclusions based on Voronoi tessellations of such data thus presents significant challenges.
Let's anticipate the error of Voronoi cell features as a function of measurement error in a simple system.
Consider a square lattice in the plane, and displace each of its points by adding an independent and identically distributed random perturbation, such as a two-dimensional Gaussian.  How do the distributions of areas and perimeters depend on the magnitude of the perturbations?
\item Repeat Problem (1) in three dimensions for a cubic lattice and consider the distributions of cell volumes and surface areas.  How do they depend on the magnitude of the random perturbations?  How do the distributions of topological features of the Voronoi cells, such as the number of edges or faces, depend on the magnitude of the perturbations?
\item Images of the Voronoi cells can be computed with a simple procedure in two dimensions. Consider a number of particles $\{\mathbf{s}_1,\mathbf{s}_2,\ldots\}\in \R^2$, and define an image covering a rectangle in $\R^2$. For each pixel location $\mathbf{x}$ in the image, assign it a different color based on computing $\arg \min_j d(\mathbf{x},\mathbf{s}_j)$ to find the nearest particle.\cite{argmin} This procedure will work for any distance metric. Try computing the Voronoi cells for different $p$-norms,\cite{heath} where $d(\mathbf{x},\mathbf{y})=\|\mathbf{x}-\mathbf{y}\|_p$, and compare their shapes to Voronoi cells with the Euclidean norm.\cite{pnorm} 
\item Instead of coloring pixels according to the nearest particle as in problem (3), try coloring particles according to the second nearest particle. What do the images look like in this case?
\item $^{\star}$ Computing Voronoi cells in arbitrary domains has several subtleties. Consider the particle arrangement in Fig.~\ref{fig:voro_bdry}(a). If the Voronoi cells are restricted to the domain $U$ shown, then one of the Voronoi cells breaks into two disconnected components [see Fig.~\ref{fig:voro_bdry}(b)]. Can criteria be placed on the shape of the domain and/or the positions of the particles to avoid this? Can the Voronoi cell definition be modified to ensure that each Voronoi cell is a single connected component?
\item $^{\star}$ Use a molecular dynamics simulator package such as LAMMPS to model a crystalline system at temperatures just below melting.  What features of the Voronoi cells indicate that the system is still crystalline?
Then heat the system until it has melted.  What features of the Voronoi cells reflect the fact that the system has melted?
\item $^{\star}$ Although icosahedral symmetry has been studied primarily in the context of many glasses, symmetric particle arrangements also appear in many other nominally disordered systems.  Even an ideal gas, for example, contains Voronoi cells that are very symmetric, including tetrahedra, cubes, and icosahedra.  Model a liquid at different temperatures and pressures. What kinds of ordered patterns, indicated by symmetric Voronoi cells, can be observed in these otherwise disordered systems? 
\item $^{\star}$ X-ray diffraction patterns are often used to characterize ordered and disordered systems.  What relations can be found between X-ray diffraction patterns and the distribution of Voronoi cell features such as areas, volumes, and topologies?
\end{enumerate}

\acknowledgments
This research was supported by a grant from the United States -- Israel Binational Science Foundation (BSF), Jerusalem, Israel through grant number 2018/170.


\begin{thebibliography}{100}
\bibitem{argmin}The notation $\arg \min_x f(x)$ is the value of $x$ for which $f(x)$ is a minimum.
\bibitem{pnorm} For a vector $\mathbf{z} = (\alpha,\beta) \in \mathbb{R}^2$, the $p$-norm is defined as $\|\mathbf{z}\|_p = (|\alpha|^p+|\beta|^p)^{1/p}$.
\end{thebibliography}
\end{document}